\documentclass[nofootinbib,twocolumn,prd,preprintnumbers,superscriptaddress,aps]{revtex4-2}

\usepackage{amsfonts}
\usepackage{graphicx}
\usepackage[colorlinks=true,linkcolor=blue,citecolor=teal,urlcolor=blue]{hyperref}
\usepackage{color}
\usepackage[dvipsnames]{xcolor}
\usepackage{amsmath}
\usepackage{braket}
\usepackage[T1]{fontenc}
\usepackage{mathrsfs}
\usepackage{enumerate}
\usepackage{bm}
\usepackage{multirow}
\usepackage{tabularx}
\usepackage[caption=false]{subfig}

\newcommand{\hodge}{{\star}}

\begin{document}

\title{Parity violation in gravitational waves and observational bounds from third-generation detectors}

\author{Matteo Califano}
\email{matteo.califano@unina.it}
\affiliation{Scuola Superiore Meridionale, Largo San Marcellino 10, 80138 Napoli, Italy}
\affiliation{Istituto Nazionale di Fisica Nucleare (INFN), Sezione di Napoli, Via Cinthia 21, 80126 Napoli, Italy}

\author{Rocco D'Agostino}
\email{rocco.dagostino@unina.it}
\affiliation{Scuola Superiore Meridionale, Largo San Marcellino 10, 80138 Napoli, Italy}
\affiliation{Istituto Nazionale di Fisica Nucleare (INFN), Sezione di Napoli, Via Cinthia 21, 80126 Napoli, Italy}

\author{Daniele Vernieri}
\email{daniele.vernieri@unina.it}
\affiliation{Dipartimento di Fisica ``E. Pancini'', Universit\`a di Napoli ``Federico II'', Via Cinthia 21, 80126 Napoli, Italy}
\affiliation{Scuola Superiore Meridionale, Largo San Marcellino 10, 80138 Napoli, Italy}
\affiliation{Istituto Nazionale di Fisica Nucleare (INFN), Sezione di Napoli, Via Cinthia 21, 80126 Napoli, Italy}

\begin{abstract}
In this paper, we analyze parity-violating effects in the propagation of gravitational waves (GWs). For this purpose, we adopt a newly proposed parametrized post-Einstenian (PPE) formalism, which encodes modified gravity corrections to the phase and amplitude of GW waveforms. In particular, we focus our study on three well-known examples of parity-violating theories, namely Chern-Simons, Symmetric Teleparallel and Hor\v ava-Lishitz gravity. For each model, we identify the PPE parameters emerging from the inclusion of parity-violating terms in the gravitational Lagrangian. Thus, we use the simulated sensitivities of third-generation GW interferometers, such as the Einstein Telescope and Cosmic Explorer, to obtain numerical bounds on the PPE coefficients and the physical parameters of binary systems. In so doing, we find that deviations from General Relativity cannot be excluded within given confidence limits. Moreover, our results show an improvement of one order of magnitude in the relative accuracy of the GW parameters compared to the values inferred from the LIGO-Virgo-KAGRA network. 
In this respect, the present work demonstrates the power of next-generation GW detectors to probe fundamental physics with unprecedented precision.
\end{abstract}

\maketitle

\preprint{ET-0337A-23}

\section{Introduction}

The gravitational wave (GW) observations by binary black holes (BBH) and/or binary neutron stars (BNS) detected by the LIGO-Virgo-KAGRA (LVK) collaboration \cite{LIGOScientific:2017zic,LIGOScientific:2018mvr,LIGOScientific:2021djp} have opened a new window to investigate fundamental physics. In this respect, the degeneracy among different theoretical scenarios brings attention to the need for investigating astrophysical sources via direct manifestations of gravitational effects. This could yield valuable physical information on the nature of gravity itself, thus playing a significant role in probing extra degrees of freedom with respect to General Relativity (GR) \cite{Ezquiaga:2017ekz,Farrugia:2018gyz,Belgacem:2018lbp,Jarv:2018bgs,LIGOScientific:2018dkp,LIGOScientific:2019fpa}.
On the other hand, the dark energy issue related to the standard cosmological model further motivated, in the last years, the search for possible extensions or modifications of GR \cite{Bengochea:2008gz,Clifton:2011jh,DAgostino:2018ngy,Nojiri:2017ncd,DAgostino:2019wko,Capozziello:2019cav,DAgostino:2020dhv,DAgostino:2022fcx}. The latter typically emerge from high-energy theories and can lead to small departures from GR in the infrared limit \cite{Stelle:1977ry,Starobinsky:1980te,Ferraro:2006jd,Deser:2007jk,Sotiriou:2008rp,Capozziello:2022rac,Bajardi:2022tzn,Capozziello:2023ccw}. 

Different impacts of modified theories of gravity on GWs can be ascribed to changes in the amplitude and/or the phase of the GW signal propagation. Changes in the phase (amplitude) may occur due to modifications of the real (imaginary) part of the dispersion relations of GWs \cite{Mirshekari:2011yq,Mewes:2019dhj,Ezquiaga:2021ler,Gong:2023ffb}.
An example of the broad class of modified gravity scenarios sharing similar consequences is represented by gravitational actions that are not invariant under a parity transformation. Parity-violating theories are characterized by an asymmetry in the propagation amplitude and speed of the left and right-handed GW polarization modes, leading to amplitude and phase birefringence, respectively \cite{Kostelecky:2016kfm,Nishizawa:2018srh,Nair:2019iur,Wang:2020pgu,Wang:2021gqm,Bombacigno:2022naf,Zhao:2022pun,Jenks:2023pmk}.

A well-known example of a parity-violating gravity scenario is the Chern-Simons (CS) theory \cite{Lue:1998mq,Alexander:2009tp,Kawai:2017kqt,Bajardi:2021hya,Sulantay:2022sag,Boudet:2022nub}, in which the Einstein-Hilbert action is extended to contain a dynamical scalar field coupled to the CS term. The parity-violating effect is due to the coupling between the (even parity) cosmological scalar field and the (odd parity) Pontryagin invariant. CS gravity takes inspiration from string theory \cite{Green:1987mn} and represents the only case of a metric theory, quadratic in the curvature and linear in the scalar field, violating parity. Moreover, the CS theory can be obtained as a limit case of the more general class of ghost-free scalar-tensor gravity \cite{Crisostomi:2017ugk,Nishizawa:2018srh,Zhao:2019xmm}, which includes parity-violating terms arising from higher-order derivatives of the scalar field.

Additional relevant examples of parity-violating theories include Symmetric Teleparallel (ST) gravity \cite{BeltranJimenez:2017tkd,Conroy:2019ibo,Capozziello:2022wgl}, which is built upon the non-metricity tensor, and some versions of Ho\v rava-Lifshitz (HL) gravity \cite{Horava:2009uw}. First introduced as a renormalizable extension of GR,  HL gravity breaks Lorentz invariance and contains higher-order derivative operators that induce parity violation \cite{Zhu:2013fja}.

A widely adopted framework to explore deviations from GR in GW propagation is provided by the parametrized post-Einstenian (PPE) formalism \cite{Yunes:2009ke}. Similarly to the post-Newtonian scheme, the PPE formalism encodes modified gravity corrections to the phase and amplitude of GR waveforms \cite{Cornish:2011ys,Huwyler:2011iq,Loutrel:2022xok}. Thus, the PPE formalism can reveal a useful tool to probe GR through GW data. Several non-PPE analyses of GW data to search for possible parity violations were previously performed only for particular waveform parametrizations \cite{Zhao:2019szi,Wang:2020cub,Okounkova:2021xjv}.
In fact, the first full PPE study of parity-violating theories has been recently presented in Ref.~\cite{Jenks:2023pmk}, where a model-independent framework was introduced to parametrize parity-violating effects in the GW-modified gravity propagation under a general scheme. 

In light of the theoretical results found in Ref.~\cite{Jenks:2023pmk}, we intend to apply the PPE framework to future GW simulated observations, in order to obtain forecast bounds on gravitational parity violation. For this purpose, we employ in our analysis the experimental sensitivities of the third-generation (3G) GW detectors, such as the Einstein Telescope (ET) \cite{Maggiore:2019uih,Branchesi:2023mws} and Cosmic Explorer (CE) \cite{Reitze:2019iox,Evans:2021gyd} interferometers. The latter have been extensively used, in recent years, to investigate scenarios beyond GR, the dark energy problem and many other fundamental questions in gravitational physics \cite{Cai:2016sby,Belgacem:2017ihm,Nishizawa:2019rra,DAgostino:2019hvh,Bonilla:2019mbm,Kalomenopoulos:2020klp,Mukherjee:2020mha,Baker:2020apq,Tasinato:2021wol,Allahyari:2021enz,Califano:2022cmo,DAgostino:2022tdk,Califano:2022syd,DAgostino:2023tgm}. 

The structure of the paper is as follows. In Sec.~\ref{sec:parity}, we introduce the parity-violating features in the GW propagation. 
In particular, we present a general parametric framework for describing parity-violating deviations from GR in terms of a few coefficients related to the modified GW amplitude and phase. Then, we take into account modifications in the GW waveform through the detector response to binary system signals. Moreover, we show how to map the PPE parameters to the parity-violating terms of modified gravity theories. 
In Sec.~\ref{sec:theories}, we consider the main theoretical frameworks where parity violation can emerge from the high-order corrections to Einstein-Hilbert action. In particular, we focus our analysis on three different scenarios: CS, ST and HL gravity models. 
In Sec.~\ref{sec:constraints}, using the simulated sensitivities of future GW detectors, we place bounds on the parity-violating coefficients and the PPE parameters of the aforementioned theories. We conclude our study in Sec.~\ref{sec:conclusions} with a discussion of the obtained results, and we draw our final considerations for future developments.

In this work, we set units such that $c=G=1$.

\section{Parity violation in the gravitational wave propagation}
\label{sec:parity}

We here show how amplitude and speed in GW propagation from BBH and BNS can be parametrized in a model-independent way. These results can be then used to probe parity violation in specific modified gravity theories. The gravitational parity-violating contribution can be encoded by a correction to the Einstein-Hilbert action:
\begin{equation}
    S=\frac{1}{2\kappa}\int d^4x\, \sqrt{-g}\, R +S_\text{PV}\,,
\end{equation}
where $\kappa\equiv 8\pi$, $g$ is the determinant of the metric tensor $g_{\mu\nu}$, and $R$ is the Ricci scalar. The term  $S_\text{PV}$ can be, in general, a function of the curvature and an auxiliary scalar field, and is responsible for modifying the GW dispersion relation.  

To study how the field equations get modified, we consider the spatially flat Fridmann-Lema\^itre-Robertson-Walker (FLRW) line element:
\begin{equation}
    ds^2=-dt^2+a^2(t)\delta_{ij}dx^i dx^j,
    \label{FLRW_background}
\end{equation}
where $a$ is the normalized scale factor as a function of cosmic time, $t$. 

Thus, we introduce linear perturbations around the background \eqref{FLRW_background}:
\begin{equation}
    ds^2=a^2(\eta)[-d\eta^2+(\delta_{ij}+h_{ij})dx^i dx^j]\,,
    \label{metric_pert}
\end{equation}
where $\eta$ is the conformal time, such that $d\eta\equiv dt/a(t)$, while $h_{ij}$ are tensor perturbations satisfying $\partial_i h^i{}_j=0=h^i{}_i$.
In particular, in this work, we focus on the two polarizations corresponding to helicity $\lambda_\text{R,L}=\pm 1$, where the subscripts \{R,\,L\} refer to the right and left-handed GW polarizations, respectively.
In the Fourier space, we can write
\begin{equation}
    h_\text{R,L}(\eta)=A_\text{R,L}(\eta)\, e^{-i[\varphi(\eta)-k_ix^i]}\,,
    \label{eq:GW_param}
\end{equation}
where $A_\text{R,L}$ is the polarization amplitude, $\varphi(\eta)$ is the GW phase and $k$ is the comoving wavenumber. 
In order to derive the GW propagation equation that violates parity, although being invariant under translations and spatial rotation, one could make use of the following assumptions:
\begin{enumerate}[(i)]
    \item deviations from GR are small, such that all modifications can be worked out within an effective field theory framework;
    \item only corrections to GR that are parity-violating are taken into account;
    \item under the assumption of locality and small deviations from GR, all modifications of Einstein's gravity are expected to be polynomial in $k$;
    \item GW wavelengths are shorter than the Universe expansion, i.e., $k\gg \mathcal{H}$, being $\mathcal{H}\equiv{a}'/a$ the conformal Hubble parameter, where the prime denotes the derivative with respect to $\eta$.
\end{enumerate}
Within the above requirements, it was shown in Ref.~\cite{Jenks:2023pmk} that the most general parametrization of parity-violating deviations in the GW propagation - including up to the second-order derivatives over time - can be expressed as 
\begin{align}
&h''_\text{R,L} + \left\{2\mathcal{H} +\lambda_\text{R,L} \sum_n k^n \left[\frac{\alpha_n \mathcal{H} }{(M_\text{PV} a)^n} + \frac{\beta_n}{(M_\text{PV} a)^{n-1}} \right]\right\}h'_\text{R,L} \nonumber   \\
&+\omega_\text{R,L}^2 h_\text{R,L} =0\,,
\label{eq:GW_prop}
\end{align}
where $\omega_\text{R,L}$ is the angular frequency,
\begin{equation}
    \omega_\text{R,L}^2=k^2\left\{1 + \lambda_\text{R,L} \sum_m k^{m-1}\left[\frac{\gamma_m \mathcal{H} }{(M_\text{PV} a)^m} + \frac{\delta_m}{(M_\text{PV} a)^{m-1}}\right]\right\},
\end{equation}
being $n=\{1,3,5,\hdots\}$ and $m=\{0,2,4,\hdots\}$. Here, $k\equiv|\vec{k}|=2\pi\nu$, where $\nu$ is the GW frequency.
In such a description, parity violation is quantified by the functions $\alpha$, $\beta$, $\gamma$ and $\delta$ depending on the conformal time, and $M_\text{PV}$ is the energy scale of the theory. It is worth noticing that modified gravity theories that violate parity usually involve dynamical scalar fields, so the expansion coefficients in the effective field framework may show a non-trivial dependence on these fields and their derivatives. 
Based on the assumption of small departures from GR, in our analysis, we consider only the leading-order corrections to GR, whose GW propagation is recovered as soon as $\alpha=\beta=\gamma=\delta=0$. 

Thus, the modified dispersion relation is obtained by replacing Eq.~\eqref{eq:GW_prop} into Eq.~\eqref{eq:GW_param}:
\begin{align}
   & \varphi'' + i\left\{2\mathcal{H} + \lambda_\text{R,L}\sum_n k^n \left[\frac{\alpha_n \mathcal{H}}{(M_\text{PV} a)^n} + \frac{\beta_n}{(M_\text{PV} a)^{n-1}} \right]\right\}\varphi' \nonumber \\
 &  + {\varphi'}^2 - \omega_\text{R,L}^2 =0\,,
    \label{eq:phi_PV}
    \end{align}
where it is assumed that the changes in the GW amplitude occur over a very long timescale compared to those relative to the phase. Considering linear perturbations around the GR background, one can write as $\varphi=\varphi_\text{GR}+\delta\varphi$, where $\delta\varphi$ accounts for amplitude and velocity birefringences in its imaginary and real parts, respectively:
\begin{equation}
    \delta\varphi = -i\lambda_\text{R,L} \delta\varphi_A + \lambda_\text{R,L} \delta\varphi_V\,.
    \label{eq:deltaphi}
\end{equation}
Consequently, a series expansion of Eq.~\eqref{eq:phi_PV} under the assumptions $\delta\varphi \ll \varphi_\text{GR}$, $\varphi'' \ll {\varphi'}^2$ and $\delta\varphi'' \ll\varphi_\text{GR}\delta\varphi'$ leads to \cite{Yunes:2010yf}
\begin{align}
\delta\varphi_A' &=  \dfrac{1}{2}\sum_n k^n\left[\frac{\alpha_n \mathcal{H}}{(M_\text{PV} a)^n } + \frac{\beta_n}{(M_\text{PV} a)^{n-1}} \right], \label{eq:deltaphiprime_A}\\
\delta\varphi_V' &= \dfrac{1}{2}\sum_m k^m\left[\frac{\gamma_m \mathcal{H}}{(M_\text{PV}  a)^m} + \frac{\delta_m}{(M_\text{PV} a)^{m-1}}\right]. \label{eq:deltaphiprime_V}
\end{align} 

The above expressions could be simplified by assuming a slow time-varying behavior for the parity-violating parameters. The latter can be thus approximated with its corresponding zeroth-order Taylor series term at the present time. Then, converting the time derivative into derivatives with respect to the redshift $z$ by means of the relation $dz/dt=-(1+z)H(z)$, integration of  Eqs.~\eqref{eq:deltaphiprime_A} and \eqref{eq:deltaphiprime_V} yields
\begin{align}
    \delta\varphi_A &= \sum_n \dfrac{k^n}{2}(1+z)^n\left[\frac{\alpha_{n_0}}{M_\text{PV} ^n}z_n + \frac{\beta_{n_0}}{M_\text{PV} ^{n-1}} D_{n+1}(z)\right], \label{eq:deltaphi_A}\\
\delta\varphi_V &=\sum_m \frac{k^m}{2}(1+z)^m\left[ \frac{\gamma_{m_0}}{M_\text{PV} ^m}z_m + \frac{\delta_{m_0}}{M_\text{PV} ^{m-1}}D_{m+1}(z)\right],
\label{eq:deltaphi_V}
\end{align}
where we made use of the following definitions \cite{Mirshekari:2011yq}:
\begin{align}
    D_\sigma(z) &= (1 +z)^{1-\sigma}\int \frac{(1 + z)^{\sigma-2}}{H(z)}dz\,, \label{eq:Dn}\\
    z_\sigma &= (1+z)^{-\sigma}\int \frac{dz}{(1 + z)^{1-\sigma}}\,.
    \label{eq:z_sigma}
\end{align}
Therefore, the modifications to the GW polarization modes can be written as
\begin{equation}
h_\text{R,L}=h^\text{(GR)}_\text{R,L}\, e^{\mp \delta\varphi_A \pm i\, \delta\varphi_V}.
\label{eq:h_modified}
\end{equation}

\subsection{Waveform modifications}

To perform a comparison with GW measurements, we shall work out the parity-violating modifications in the standard $+/\times$ basis. Specifically, from the circular polarization modes, one can define the linear modes 
\begin{equation}
h_{+} = \frac{h_R + h_L}{\sqrt{2}}\,, \quad h_\times = i \frac{h_R - h_L}{\sqrt{2}}\,.
\end{equation}
Thus, expanding Eq.~\eqref{eq:h_modified} at the first order gives
\begin{align}
h_+ &= h^\text{(GR)}_+ - i\, \delta\varphi_A h^\text{(GR)}_\times + \delta\varphi_V h^\text{(GR)}_\times, \label{eq:h_plus}\\
h_\times &= h^\text{(GR)}_\times + i\, \delta\varphi_A h^\text{(GR)}_+ - \delta\varphi_V h^\text{(GR)}_+ \label{eq:h_cross}\,.
\end{align}

For a given detector, the measured GW response function may be written as
\begin{equation}
    \tilde{h} = F_+\tilde{h}_+ + F_\times \tilde{h}_\times\,,
    \label{eq:h_tilde}
\end{equation}
where the beam functions $F_{+,\times}$ depend on the polarization angle and location of the GW source in the sky \cite{Sathyaprakash:2009xs}. 
In the PN approximation, we can write the GR polarization modes in the case of quasi-circular and non-precessing binaries as \cite{Damour:2004bz}
\begin{align}
    \tilde{h}^\text{(GR)}_{+} = A(1 + \xi^2) e^{i\psi}\,, \label{eq:h_tilde_plus}\\
    \tilde{h}^\text{(GR)}_\times = 2A\xi e^{i(\psi + \pi/2)}\,, \label{eq:h_tilde_cross}
\end{align}
where $A$ and $\psi$ are the GW amplitude and phase, respectively, in the stationary phase regime. Moreover, $\xi\equiv \cos \iota$, being $\iota$ the inclination angle between the line of sight and the angular momentum vector of the source.
The detector response as a function of the GW frequency is given by
\begin{equation}
    \tilde{h}_\text{GR}(\nu) = A\nu^{-7/6}e^{i(\psi + \delta\psi)},
\end{equation}
where 
\begin{align}
    A & = \sqrt{\frac{5}{96\pi^{4/3}}}\frac{\mathcal{M}^{5/6}}{d_L(z)}\sqrt{F_+^2(1+ \xi^2)^2 + 4 F_\times^2 \xi^2}\,, \\
    \delta\psi & =\tan^{-1}\left[\frac{2F_\times\xi}{F_+(1+\xi^2)}\right]\,, 
\end{align}
being $\mathcal{M}\equiv (m_1m_2)^{3/5}\times (m_1+m_2)^{-1/5}$ the chirp mass of the binary system composed by the objects with masses $m_1$ and $m_2$, and $d_L(z)$ the luminosity distance\footnote{Following the prescription of Eq.~\eqref{eq:Dn}, $d_L(z)=(1+z)^2 D_2(z)$, where $D_2(z)$ coincides with the angular diameter distance.}.

Hence, one can feature the parity-violating GW propagation as
\begin{equation}
    \tilde{h} = \tilde{h}_\text{GR}(1 + \delta A_A + \delta A_V)e^{i(\delta\psi_A + \delta\psi_V)}\,.
    \label{eq:h_tilde2}
\end{equation}
The corrections $\delta\psi_A$ and $\delta\psi_V$ are found by plugging Eqs.~\eqref{eq:h_plus} and \eqref{eq:h_cross} into Eq.~\eqref{eq:h_tilde}, and then expanding the resulting expressions for the amplitude and phase at the linear order in $\delta \varphi_{A,V}$. In doing so, we obtain
\begin{align} 
& \delta A_A + \delta A_V =   f(F_{+,\times}, \xi)\delta\varphi_A -  g(F_{+,\times}, \xi)\delta\varphi_V \,,  \label{eq:delta_A} \\
& \delta\psi_A + \delta\psi_V =  g(F_{+,\times}, \xi)\delta\varphi_A   +  f(F_{+,\times}, \xi)\delta\varphi_V\,, \label{eq:delta_Psi}
\end{align}
where we introduced the following auxiliary functions:
\begin{align}
   f(F_{+,\times}, \xi) &:= \frac{ 2(F_+^2 + F_\times^2)(1 + \xi^2)\xi}{4F_\times^2\xi^2 + F_+^2(1 + \xi^2)^2}\,, \\
   g(F_{+,\times}, \xi) &:= \frac{F_+F_\times(1-\xi^2)^2}{4F_\times^2\xi^2 \label{g}
+ F_+^2(1 + \xi^2)^2}\,.
\end{align}
In view Eqs.~\eqref{eq:delta_A} and \eqref{eq:delta_Psi}, Eq.~\eqref{eq:h_tilde2} finally becomes
\begin{align} 
\tilde{h} =&\ \tilde{h}_\text{GR}\left[1 + f(F_{+,\times}, \xi)\delta\varphi_A - g(F_{+,\times}, \xi)\delta\varphi_V \right]\nonumber \\
&\times \exp\left\{i\left[g(F_{+,\times}, \xi)\delta\varphi_A + f(F_{+,\times}, \xi)\delta\varphi_V\right]\right\}.
\label{eq:h_tilde3}
\end{align} 

\subsection{PPE formalism}

At this point, we shall show how the parity-violating modifications in the propagation of GWs can be framed within the PPE formalism \cite{Yunes:2009ke,Jenks:2023pmk}. For this purpose, let us consider the following PPE waveform:
\begin{equation}
    \tilde{h}_\text{PPE} = \tilde{h}_\text{GR}\left(1 + \alpha_\text{PPE} u^{a_\text{PPE}} \right)\exp\left\{i\, \beta_\text{PPE} u^{b_\text{PPE}}\right\}\,.
\end{equation}
Here, the parameters $a_\text{PPE}$, $\alpha_\text{PPE}$, $\beta_\text{PPE}$ and $b_\text{PPE}$ are dimensionless coefficients to be mapped to different gravity models, and $u=\pi\nu\mathcal{M}$.  

Then, to account for the parity-violating theories, we can use Eq.~\eqref{eq:h_tilde3} with the explicit forms of $\delta\varphi_A$ and $\delta\varphi_V$. In this way, one finds the mapping
\begin{equation}\label{eq: ppE}
    \tilde{h} = \tilde{h}_\text{GR} \left(1+\sum_{a_\text{PPE}} u^{a_\text{PPE}}\alpha_{a_\text{PPE}}^{\text{(PPE)}}\right)\exp\left\{i\sum_{b_\text{PPE}}u^{b_\text{PPE}}\beta_{b_\text{PPE}}^{\text{(PPE)}}\right\}
\end{equation}
from which we infer $a_\text{PPE}=b_\text{PPE}=(n,m)$. Specifically, for $a_\text{PPE}=b_\text{PPE}=n$, we have\footnote{Notice that $u=\mathcal{M}k/2$\,.}
\begin{align}
\alpha^\text{(PPE)}_{n}  &= \left[\frac{2(1+z)}{\mathcal{M}M_\text{PV}}\right]^{n}\frac{f(F_{+,\times},\xi)}{2} \big[\alpha_{n_0}z_n + M_\text{PV} \beta_{n_0}D_{n+1}(z)\big], \label{eq:alphaPPE_n}\\
\beta^\text{(PPE)}_{n} &= \left[\frac{2(1+z)}{\mathcal{M}M_\text{PV}}\right]^{n}\frac{g(F_{+,\times},\xi)}{2}\big[\alpha_{n_0}z_n+ M_\text{PV}\beta_{n_0}D_{n+1}(z)\big]. \label{eq:betaPPE_n}
\end{align}
On the other hand, for $a_\text{PPE}=b_\text{PPE}=m$, one has
\begin{align}
\alpha^\text{(PPE)}_{m} &=- \left[\frac{2(1+z)}{\mathcal{M}M_\text{PV} }\right]^{m}\frac{g(F_{+,\times},\xi)}{2} \big[\gamma_{m_0}z_m + M_\text{PV}\delta_{m_0}D_{m+1}(z)\big],\\
\beta^\text{(PPE)}_{m} &= \left[\frac{2(1+z)}{\mathcal{M}M_\text{PV} }\right]^{m}\frac{f(F_{+,\times},\xi)}{2} \big[\gamma_{m_0}z_m + M_\text{PV} \delta_{m_0}D_{m+1}\big].
\end{align}

Furthermore, it is possible to frame the GW linear polarization modes within the PPE formalism. In particular, we parametrize the detector response as
\begin{align}
    \tilde{h}_+ &= \tilde{h}^\text{(GR)}_+(1 + \delta A_+)e^{i\delta\psi_+}\,,\\
    \tilde{h}_\times &= \tilde{h}^\text{(GR)}_\times(1 + \delta A_\times)e^{i\delta\psi_\times}\,,
\end{align}
that can be combined with Eqs.~\eqref{eq:h_tilde_plus} and \eqref{eq:h_tilde_cross} to obtain
\begin{align}
\tilde{h}_{+,\times} &= \tilde{h}^\text{(GR)}_{+,\times}\Big[1 + \zeta_{+,\times}(\xi)\delta\varphi_A \Big]e^{i\zeta_{+,\times}(\xi)\,\delta\varphi_V}\,,
\end{align}
where we introduced
\begin{equation}
    \zeta_+(\xi):=\dfrac{2\xi}{1+\xi^2}\ , \quad \zeta_\times(\xi):=\dfrac{(1+\xi)^2}{2\xi}\,.
\end{equation}
Then, in this case, the PPE parameters are $a_\text{PPE} = n\,, \ b_\text{PPE}= m$ and
\begin{align}
&\alpha^\text{(PPE)}_{n}  = \left( \frac{2}{\mathcal{M}M_\text{PV}}\right)^n \frac{\zeta_{+,\times}(\xi)}{2} \big[\alpha_{n_0}z_n + M_\text{PV}\beta_{n_0} D_{n+1}(z)\big],\\
&\beta^\text{(PPE)}_{m} =\left(\frac{2}{\mathcal{M}M_\text{PV} }\right)^m \frac{\zeta_{+,\times}(\xi)}{2}\big[\gamma_{m_0}z_m+ M_\text{PV}\delta_{m_0} D_{m+1}(z)\big]. 
\end{align}

\section{Parity-violating theories of gravity}
\label{sec:theories}

In this Section, we briefly describe the main features of the most relevant parity-violating modified gravity theories. Thus, we infer the expressions of the PPE parameters for the specific model under consideration.

\subsection{Chern-Simons gravity}

As mentioned earlier, the CS theory is one of the most well-studied scenarios leading to parity violation \cite{Jackiw:2003pm}. In this case, the modified gravity action is given by
\begin{equation}
    S_\text{CS} = \dfrac{1}{2\kappa}\int d^4 x\, \sqrt{-g} \left(R+\frac{\alpha_\text{CS}}{4}\vartheta R\, ^\hodge R\right), 
    \label{action_CS}
\end{equation}
where $\alpha_\text{CS}$ is a coupling constant, $\vartheta$ is a dynamical scalar field, and $R\, ^\hodge R$ is the Pontryagin density defined as
\begin{equation}
R\, ^\hodge R=\frac{1}{2}R_{abcd}\,\varepsilon^{abef}R^{cd}{}_{ef}\,, 
\end{equation}
where $\varepsilon^{abcd}$ is the Levi-Civita tensor.

Considering linear perturbations as in Eq.~\eqref{metric_pert}, the equations of motion for the tensor modes are given by (see \cite{Alexander:2004wk} for the details) 
\begin{equation}
     \mathcal{D}^j{}_i + \frac{\varepsilon^{sjl}}{a^2}\big[(\vartheta'' - 2\mathcal{H}\vartheta')\partial_s h_{il}' + \vartheta' \partial_s \mathcal{D}_{il}\big]=0\,,
\end{equation}
where we defined  
\begin{equation}
    \mathcal{D}_{ij}:=h_{ij}''+2\mathcal{H}h_{ij}'-\partial_l\partial^lh_{ij}\,.
\end{equation}
Moreover, when searching for plane-wave solutions, the GW polarization modes obey the dispersion relation \cite{Yunes:2010yf}
\begin{equation}
i\ddot{\varphi}+\dot{\varphi}^2 - k^2 = -i\dfrac{k\lambda_\text{R,L}\alpha_\text{CS}\ddot{\vartheta}}{1-k\lambda_\text{R,L}\alpha_\text{CS}\dot{\vartheta}}\dot{\varphi}\,.
\label{eq:disp_CS}
\end{equation}
Then, making use of the equation of motion for the scalar field, $\ddot{\vartheta}+2H\dot{\vartheta}=0$, and linearizing Eq.~\eqref{eq:disp_CS}, one finally obtains
\begin{equation}
    \delta\varphi = -2ik\lambda_\text{R,L} \alpha_{\text{CS}_0} \dot{\vartheta}_0 z\,.
    \label{eq:deltaphi_CS}
\end{equation}
Since the units of the $\alpha_\text{CS}\dot{\vartheta}_0$ term are those of a length, we operate the redefinition $\alpha_\text{CS}\rightarrow \tilde{\alpha}_\text{CS}=\alpha_\text{CS}M_\text{PV}$ in order for Eq.~\eqref{eq:deltaphi_CS} to be dimensionless.

Now, if we compare Eq.~\eqref{eq:deltaphi_CS} to Eq.~\eqref{eq:deltaphi} with the help of the expressions \eqref{eq:deltaphi_A} and \eqref{eq:deltaphi_V}, we infer\footnote{From Eq.~\eqref{eq:z_sigma}, one finds $z_1=z(1+z)^{-1}$\,.} $\alpha_1=4\tilde{\alpha}_\text{CS}\dot{\vartheta}$, whereas all the other parity-violating coefficients are vanishing. 
Thus, from Eqs.~\eqref{eq:alphaPPE_n} and \eqref{eq:betaPPE_n}, we obtain the PPE parameters corresponding to the CS theory:
\begin{align}
\label{eq: alpha cs}
\alpha_1^\text{(PPE)}&=\frac{f(F_{+,\times},\xi)}{\mathcal{M}M_\text{PV}}\alpha_{1_0} z\,, \\
\beta_1^\text{(PPE)}&=\frac{g(F_{+,\times},\xi)}{\mathcal{M}M_\text{PV}}\alpha_{1_0} z\,. \label{eq: beta cs}
\end{align}

\subsection{Symmetric Teleparallel gravity}

Another relevant parity-violating theory we take into account in our study is ST gravity. In particular, the ST Equivalent to GR action is given as \cite{Nester:1998mp}
\begin{equation}
    S_\text{STEGR}=-\frac{1}{2\kappa}\int d^4x\, \sqrt{-g}\, \mathcal{L}_\text{STEGR}\,,
\end{equation}
where
\begin{equation}
\mathcal{L}_\text{STEGR}=-\frac{1}{4}Q_{abc}Q^{abc}+ \frac{1}{2}Q_{abc}Q^{bac}+ \frac{1}{4}Q_aQ^a - \frac{1}{2}Q_a\tilde{Q}^a\,.
\end{equation}
Here, $Q_{abc}\equiv \nabla_a g_{bc}$ is the non-metricity tensor, whose contractions obey the relations
\begin{equation}
    Q_a = g^{bc}Q_{abc}\,, \quad \tilde{Q}_c = g^{ab}Q_{abc}\,.
\end{equation}

In ST geometry with coupling to a scalar field $\phi$, once introducing perturbations as in Eq.~\eqref{metric_pert}, the only non-vanishing parity-violating Lagrangians that are second-order in derivatives are (see Ref.~\cite{Conroy:2019ibo} for the details)
    \begin{align}
L_\text{PV,1}^{(2)}&= \varepsilon^{abcd}\partial_c\phi\, \partial^f\phi\, Q_{abe}Q_{fd}{}^e\,,\\
L_\text{PV,2}^{(2)} &= \varepsilon^{abcd}\partial_f\phi\,\partial^f\phi\, Q_{abe}Q_{cd}{}^e\,.
\end{align}
Hence, the parity-violation action can be written as
\begin{equation}
    S_\text{PV}^{(2)}=\dfrac{1}{2\kappa}\int d^4x\, \sqrt{-g}\, \alpha_{\text{ST},i} L_{\text{PV},i}^{(2)}\,,
\end{equation}
where $\alpha_{\text{ST},i}\,(i=1,2)$ are arbitrary function of $\phi$ and the related kinetic term. One can show that the two Lagrangians actually differ only by a constant and, thus, the ST modified gravity action may be written as
\begin{equation}
    S_\text{ST}^{(2)}=\dfrac{1}{2\kappa}\int d^4x\, a^3\left(\mathcal{L}_\text{STEGR}^{(2)}+\mathcal{L}_\text{PV}^{(2)}\right),
    \label{eq:action_ST_2}
\end{equation}
with 
\begin{align}   \mathcal{L}_\text{STEGR}^{(2)}&=\frac{1}{4}\left(\dot{h}^{ij}\dot{h}_{ij}-\partial^k h_{ij}\partial_k h^{ij}\right),\\
    \mathcal{L}_\text{PV}^{(2)}&=\frac{H}{a}\alpha_\text{ST}\, \varepsilon^{ijk} h_k{}^l\partial_i h_{jl}\,,
\end{align}
where $\alpha_\text{ST}$ can be thought of as a generic function of time.
Then, the equations of motion for tensor perturbations are given by
\begin{equation}
    h_{ij}'' + 2\mathcal{H} h_{ij}' - \partial^2 h_{ij} - 4\mathcal{H}\alpha_\text{ST}\,\varepsilon_{kl(i}\partial^k h^l_{j)} = 0\,,
    \label{eq:eom_ST}
\end{equation}
where $\partial^2\equiv \delta_{ij}\partial^i\partial^j$.
Then, the dispersion relation reads \cite{Jenks:2023pmk}
\begin{equation}
    i\varphi'' + 2i\mathcal{H}\varphi' + {\varphi'}^2 - k^2 + 4 k\mathcal{H} \alpha_\text{ST}\lambda_\text{R,L} = 0\,,
\end{equation}
and one finds
\begin{equation}
    \delta\varphi = -2\lambda_\text{R,L} \alpha_{\text{ST}_0} \ln(1+z)\,.
\end{equation}
From the comparison between the latter\footnote{Notice that, according to Eq.~\eqref{eq:z_sigma}, $\ln(1+z)=z_0$.} and Eq.~{\eqref{eq:deltaphi}}, we can map $\gamma_{0}=-4\alpha_\text{ST}$, while all the other parity-violating coefficients are zero. Moreover, the PPE parameters read 
\begin{align}
    \alpha_0^\text{(PPE)} & = -\frac{z_0\gamma_{0_0}}{2}g(F_{+,\times},\xi)\,, \\
    \beta_0^\text{(PPE)} & = \frac{z_0\gamma_{0_0}}{2}f(F_{+,\times},\xi)\,. 
\end{align}

Furthermore, if one considers the third-order terms in derivatives, the non-vanishing parity-violating Lagrangians are \cite{Conroy:2019ibo}
\begin{align}
    L^{(3)}_\text{PV,1}&=\varepsilon^{abcd}\partial_d\phi\nabla_a Q_{fb}{}^e Q^f{}_{ce}\,, \\
    L^{(3)}_\text{PV,2}&=\varepsilon^{abcd}\partial_d\phi\nabla_f Q^{fb}{}_a Q_{bce}\,, \\
    L^{(3)}_\text{PV,3}&=\varepsilon^{abcd}\partial^e\phi\nabla_a Q_{ebf} Q_{cd}{}^f\,.
\end{align}
In this case, the modified gravity action for GW propagation may be written as
\begin{align}
    S_\text{ST}^{(3)}= \frac{1}{2\kappa} \int d^4 x\, a^3 & \left(\frac{\beta_1}{a^3M_\text{PV}}\mathcal{L}_\text{PV,1}^{(3)} + \frac{\beta_2}{aM_\text{PV}}\mathcal{L}_\text{PV,2}^{(3)}\right. \nonumber \\
    &\quad \left.+ \frac{\beta_3}{aM_\text{PV}}\mathcal{L}_\text{PV,3}^{(3)}\right),
\end{align}
where $\beta_{i}\,(i=1,2,3)$ are generic time-dependent functions, and
\begin{align}
\mathcal{L}_\text{PV,1}^{(3)} &= \varepsilon^{ijk}\partial^2 h_j{}^l \partial_i h_{kl}\,,\\
    \mathcal{L}_\text{PV,2}^{(3)} &= 2H\varepsilon^{ijk} \dot{h}_j{}^l\partial_i h_{kl}\,,\\
    \mathcal{L}_\text{PV,3}^{(3)} &= \varepsilon^{ijk}\dot{h}_j{}^l \partial_i\dot{h}_{kl}\,.
\end{align}
Thus, the equations of motion read
\begin{align}
   & h_{ij}'' + 2\mathcal{H} h_{ij}' - \partial^2 h_{ij}-\dfrac{\varepsilon_{(ilk}}{aM_\text{PV}}\Big[(-\beta_1\partial^2  +\tilde{\beta}_1)\partial^l h^k{}_{j)}\nonumber \\
    &   +\tilde{\beta}_2 \bar{g}_{j)q}\partial^l h^{\prime kq}+\beta_3 \bar{g}_{j)q}\partial^l h^{\prime\prime kq} \Big],
\end{align}
where $\bar{g}_{ij}$ is the metric tensor for the line element \eqref{FLRW_background}, and we defined 
\begin{align}
    \tilde{\beta}_1 & \equiv \beta_2'\mathcal{H}+\beta_2\mathcal{H}'+3(\beta_3'\mathcal{H}+\beta_3\mathcal{H}')+\beta_3 \mathcal{H}^2\,, \\
    \tilde{\beta}_2 & \equiv \beta_2'+3\beta_2\mathcal{H}  \,.
\end{align}
The dispersion relation is given by \cite{Jenks:2023pmk}
\begin{align}
    &\varphi'' + {\varphi'}^2 + i\left[2\mathcal{H}  + \frac{ \lambda_\text{R,L} k \mathcal{H}}{aM_\text{PV}}\left(3\beta_2 - 2\beta_3\right)+ \frac{ \lambda_\text{R,L}  k}{aM_\text{PV}}\beta_2'\right]\varphi'  \nonumber \\
    &- k^2\left[1 +  \frac{\lambda_\text{R,L} k}{aM_\text{PV}}\left(\beta_1- \beta_3\right)\right] =0\,,
\end{align}
and we have
\begin{align}
    \delta\varphi =& -\frac{i\lambda_\text{R,L} k}{2M_\text{PV}}\left[\dot{\beta}_{2_0}(1+z)D_2(z) + 3({\beta_2}_0 - {\beta_3}_0 )z\right] \nonumber \\
    & + \frac{\lambda_\text{R,L}k^2 }{2M_\text{PV}}({\beta_1}_0 - {\beta_2}_0)(1+z)^2 D_3(z)\,. 
\end{align}
Therefore, by rescaling $\dot{\beta}_2 \rightarrow \dot{\tilde{\beta}}_2 = \dot{\beta}_2 /M_\text{PV}$, we find that the non-vanishing parity-violating coefficients are
\begin{align}
\alpha_1 & =3(\beta_2-\beta_3)\,, \\
\beta_1 & =\dot{\tilde{\beta}}_2\,, \\
\delta_2 & =\beta_1-\beta_2\,.
\end{align}
As far as the PPE coefficients are concerned, we find
\begin{align}
    \alpha_1^\text{(PPE)} & =\dfrac{f(F_{+,\times},\xi)}{2}\left[\frac{\alpha_{1_0}}{M_\text{PV}}z+\beta_{1_0}(1+z)D_2(z)\right],     \label{eq: a 1 ppe stegr}\\
    \beta_1^\text{(PPE)}&= \dfrac{g(F_{+,\times},\xi)}{2}\left[\frac{\alpha_{1_0}}{M_\text{PV}}z+\beta_{1_0}(1+z)D_2(z)\right], \label{eq: b 1 ppe stegr}\\
    \alpha_2^\text{(PPE)} &= -\dfrac{2g(F_{+,\times},\xi)}{\mathcal{M}^2M_\text{PV}}\delta_{2_0}(1+z)^2 D_3(z)\,, \label{eq: a 2 ppe stegr}\\
    \beta_2^\text{(PPE)}& = \dfrac{2f(F_{+,\times},\xi)}{\mathcal{M}^2M_\text{PV}}\delta_{2_0}(1+z)^2 D_3(z)\,. \label{eq: b 2 ppe stegr}
\end{align}

\subsection{Ho\v rava-Lifshitz gravity}

The HL theory of gravity was first proposed in Ref.~\cite{Horava:2009uw}, where it was shown that both Lorentz symmetry breaking and parity violation can occur. The most general form of the gravitational part of the HL action that is invariant under parity transformations is given by \cite{Zhu:2011xe,Zhu:2011yu} 
\begin{equation}
    S_\text{HL}=\dfrac{1}{2\kappa}\int d^4x \sqrt{-g}\,N\left(\mathcal{L}_K-\mathcal{L}_V^{(R)}-\mathcal{L}_V^{(a)}+\mathcal{L}_A+\mathcal{L}_\phi\right)
    \label{eq:action_HL}
\end{equation}
where
\begin{align}
\mathcal{L}_K&=\,K_{ij}K^{ij}-\lambda K^2\,, \\
\mathcal{L}_V^{(R)}&=\dfrac{g_0}{2\kappa}+g_1 R+2\kappa\left(g_2 R^2+g_3 R_{ij}R^{ij}\right)+4\kappa^2g_5 C_{ij}C^{ij}\,, \\
\mathcal{L}_V^{(a)}&=-\xi_0a_ia^i+2\kappa\left[\xi_1(a_i a^i)^2+\xi_2(a^i{}_i)^2+\xi_3(a_ia^i)a^j{}_j \right. \nonumber \\
&\quad +\left.\xi_4a^{ij}a_{ij}+\xi_5 a_i a^i R+\xi_6a_ia_jR^{ij}+\xi_7a^i{}_iR\right]\nonumber \\
&\quad +4\kappa^2\xi_8(\nabla^2a^i)^2\,, \\
\mathcal{L}_A&=\frac{A}{N}(2\Lambda-R)\,, \\
\mathcal{L}_\phi&=\phi G^{ij}(2K_{ij}+\nabla_i\nabla_j\phi+a_i\nabla_j\phi)+(1-\lambda)\times  \nonumber \\
&\quad \times \left[(\nabla^2\phi+a_i\nabla^i\phi)^2+2\left(\nabla^2\phi+a_i\nabla^i\phi\right)K\right] \nonumber \\
&\quad + \frac{1}{3}\mathcal{G}^{ijlk}\Big[4(\nabla_i\nabla_j\phi)a_{(k}\nabla_{l)}\phi+5(a_{(i}\nabla_{(j}\phi)a_{(k}\nabla_{l)}\phi \nonumber \\
&\quad +2\left(\nabla_{(i}\phi\right)a_{j)(k}\nabla_{l)}\phi+6K_{ij}a_{l(}\nabla_{k)}\phi\Big].
\end{align}
Here, $K_{ij}$ and $C_{ij}$ are the extrinsic curvature and the Cotton tensor, respectively, defined as
\begin{align}
    K_{ij}&:=\dfrac{1}{2N}\left(-\dot{g}_{ij}+\nabla_i N_j+\nabla_j N_i\right), \\
    C^{ij}&:=\frac{\varepsilon^{ikl}}{\sqrt{g}}\nabla_k\left(R^j_l-\frac{1}{4}R\delta_l^j\right),
\end{align}
while $\lambda$, $g_i\,(0=2,\hdots,5)$ and $\xi_i\,(i=0,\hdots8)$ are coupling constants.
Also, $a_i=\partial_i(\ln N)$, $a_{ij}=\nabla_j a_i$, being $N_i=g_{ij}N^j$ the shift vector and $N$ the lapse function in the Arnowitt-Deser-Misner decomposition. 
Moreover, $A$ and $\phi$ are the $U(1)$ gauge field and  Newtonian prepotential, respectively, whereas 
$\mathcal{G}^{ijlk}\equiv g^{il}g^{jk}-g^{ij}g^{kl}$,
and $G_{ij}$ is the Einstein tensor including the contribution of the cosmological constant, $\Lambda$:
\begin{equation}
    G_{ij}:=R_{ij}-\frac{1}{2}g_{ij}R+g_{ij}\Lambda\,.
\end{equation}
 The parity-violating effects can be studied by including in the action \eqref{eq:action_HL} the fifth and sixth-order spatial derivative operators \cite{Zhu:2013fja}:
\begin{align}
\mathcal{L}_\text{PV}=\dfrac{\alpha_\text{HL,0}}{M_\text{PV}^3}K_{ij}R^{ij}+\dfrac{\alpha_\text{HL,1}}{M_\text{PV}}\omega_3(\Gamma)+\dfrac{\alpha_\text{HL,2}}{M_\text{PV}^3}\varepsilon^{ijk}R_{il}\nabla^2_j R^l_k\,,
    \label{eq:L_PV_HL}
\end{align}
where $\alpha_{\text{HL},i}\,(i=0,1,2)$ are dimensionless  constants, and $\omega_3(\Gamma)$ is the three-dimensional CS term:
\begin{equation}
\omega_3(\Gamma):=\dfrac{\varepsilon^{ijk}}{\sqrt{-g}}\left( \Gamma_{jl}^m\partial_i\Gamma_{km}^l+\frac{2}{3}\Gamma_{il}^n\Gamma_{jm}^l\Gamma_{kn}^m\right).
\end{equation}
It is worth remarking that, in Eq.~\eqref{eq:action_HL}, we neglected extra fifth-order operators that do not contribute to the tensor perturbations.

Assuming the metric \eqref{metric_pert} under the gauge $\phi=0$, one has $N=a(\eta)$ and $N^i=A=0$ \cite{Zhu:2011yu}. 
Then, considering up to the second-order derivatives of the tensor perturbations, the field equations read
\begin{align}
    &h_{ij}'' + 2\mathcal{H} h_{ij}' -\alpha_\text{HL}^2\partial^2 h_{ij} \nonumber \\ 
    &+ \varepsilon_i{}^{lk}\left[\frac{2\alpha_\text{HL,1}}{M_\text{PV}a} + \frac{\alpha_\text{HL,2}}{(M_\text{PV}a)^3}\partial^2\right]\partial_l(\partial^2 h_{jk}) = 0\,,
\end{align}
where $\alpha_\text{HL}^2\equiv 1+3\alpha_\text{HL,0}\mathcal{H}/(2M_\text{PV}^3a)$.
We notice that a healthy behavior of the theory on infrared scales requires $\alpha_\text{HL}^2\simeq 1$. This implies that one can set $\alpha_\text{HL,0}=0$ without any loss of generality.
In this case, the GW dispersion relation can be written as \cite{Jenks:2023pmk}
\begin{equation}
    i\varphi'' + \varphi'^2 + 2 i \mathcal{H}\varphi' - k^2  + \lambda_\text{R,L} \left[\frac{2\alpha_\text{HL,1}}{M_\text{PV}a} - \frac{\alpha_\text{HL,2}k^2}{(M_\text{PV}a)^3}\right]k^3=0\,,
\end{equation}
which leads to
\begin{align}
    \delta\varphi =& -\frac{\alpha_{\text{HL,1}_0}\lambda_\text{R,L}}{2M_\text{PV}}k^2 (1+z)^2 D_3(z)  \nonumber \\
   & + \frac{\alpha_{\text{HL,2}_0}\lambda_\text{R,L} }{2M_\text{PV}^3}k^4 (1+z)^4 D_5(z)\,.
\end{align}
Comparing the latter with the general parametrization framework given in Eq.~\eqref{eq:deltaphi}, we find the non-zero parity-violating coefficients to be
\begin{equation}
    \delta_2=-\alpha_\text{HL,1}\,, \quad \delta_4=\alpha_\text{HL,2}\,.
\end{equation}
Then, the PPE parameters are obtained as
\begin{align}
    \alpha_2^\text{(PPE)}&=-\frac{2g(F_{+,\times},\xi)}{\mathcal{M}^2M_\text{PV}}\delta_{2_0}(1+z)^2 D_3(z)\,, \label{eq: a 2 HL}\\
    \beta_2^\text{(PPE)}&= \frac{2f(F_{+,\times},\xi)}{\mathcal{M}^2M_\text{PV}}\delta_{2_0}(1+z)^2 D_3(z)\,, \label{eq: b 2 HL}\\
    \alpha_4^\text{(PPE)}&=-\frac{8g(F_{+,\times},\xi)}{\mathcal{M}^4M_\text{PV}^3}\delta_{4_0}(1+z)^4 D_5(z)\,, \label{eq: a 4 HL}\\
    \beta_4^\text{(PPE)}&=\frac{8f(F_{+,\times},\xi)}{\mathcal{M}^4M_\text{PV}^3}\delta_{4_0}(1+z)^4 D_5(z)\,.\label{eq: b 4 HL}
\end{align}

\begin{figure}
  \centering
\includegraphics[width=0.45\textwidth]{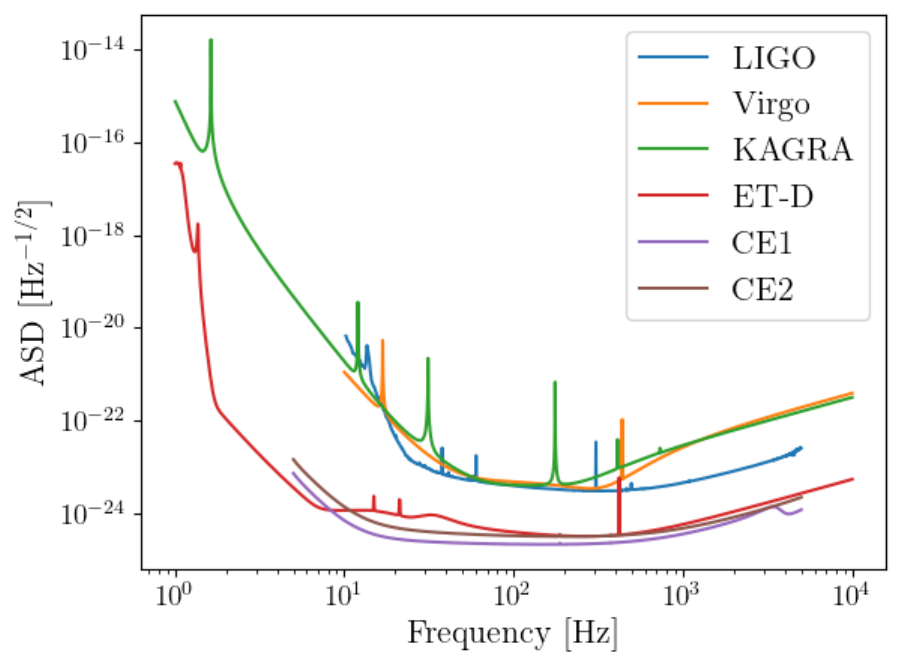}
    \caption{Amplitude spectral density for 2G and 3G detectors.}
    \label{fig: asd}
\end{figure}

\begin{table}
    \setlength{\tabcolsep}{0.35em}
     \renewcommand{\arraystretch}{1.3}
    \begin{tabular}{c c c c c c }
    \hline
    Detector & Latitude &Longitude& \shortstack{x-arm\\ azimuth} & \shortstack{y-arm\\  azimuth}& $f_{ini}\, [\text{Hz}]$\\
    \hline
    ET-1 & 0.7615 &  0.1833 & 0.3392 &5.5752& 1\\
    ET-2 & 0.7629 & 0.1841 & 4.5280 & 3.4808 & 1 \\
    ET-3 & 0.7627 & 0.1819 & 2.4336& 1.3864 & 1\\
    CE1 & 0.7613 & $-2.0281$ &1.5708 & 0 & 5 \\
    CE2 & $-0.5811$ & 2.6021 &2.3562 & 0.7854& 5 \\
    \hline
    \end{tabular}
    \caption{Localization and the power spectral density lowest frequency of the detectors considered in this study.}
    \label{tab:detectors}
\end{table} 

\begin{figure*}
    \centering
\includegraphics[width=0.8\textwidth]{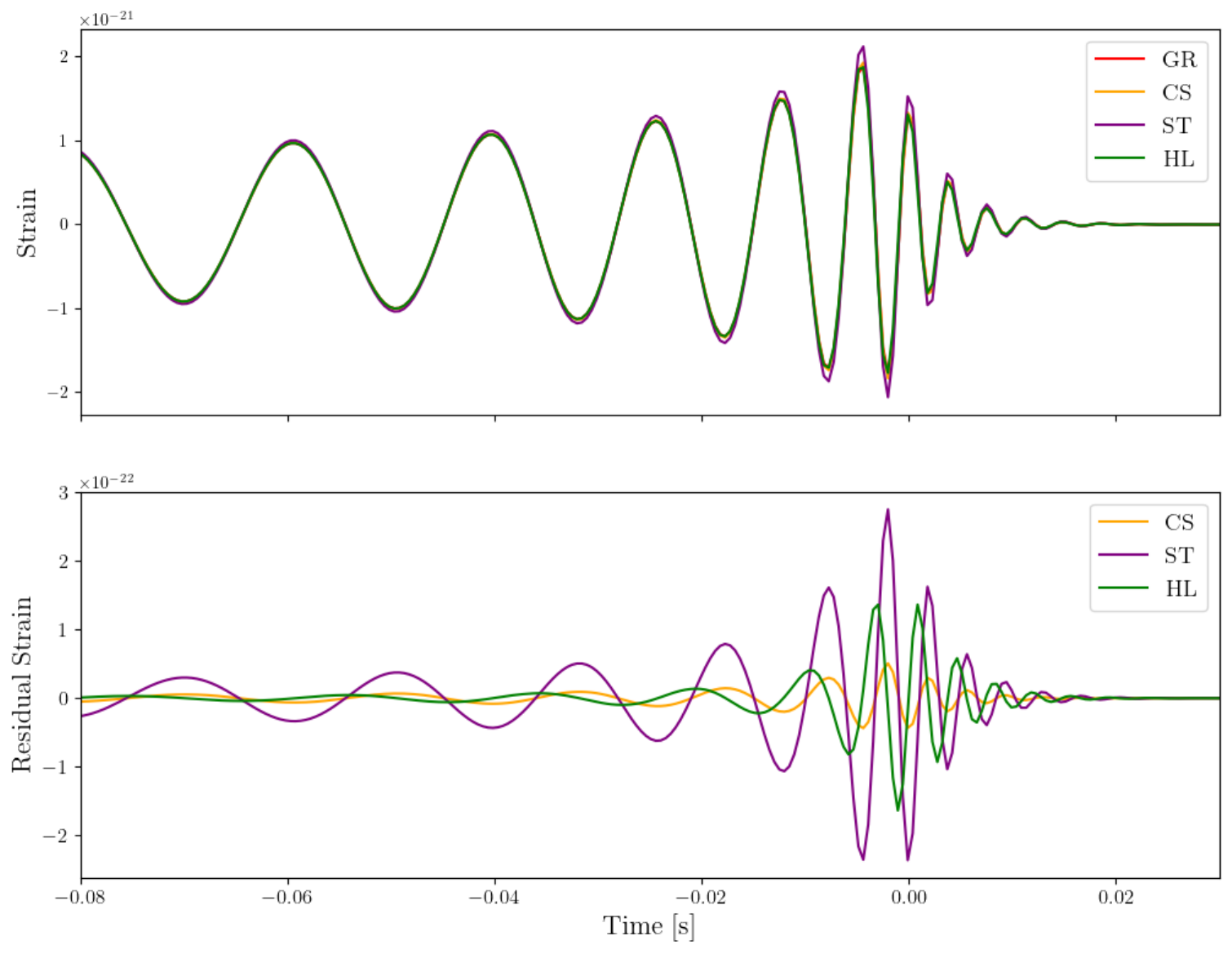}
    \caption{Top panel: waveforms for a GW150914-like event for different values of the $\bm{\theta}_\text{PPE}$ parameters. Bottom panel: residual strain, i.e., $\tilde{h}_\text{GR}- \tilde{h}_\text{PPE}$, for the different theories under consideration in this work.}
    \label{fig: strain}
\end{figure*}

\section{Observational constraints}
\label{sec:constraints}

In this Section, we study the power constraint of 3G detectors on parity-violating theories. In particular, we focus on the capabilities of ET and CE.
For ET, we consider a triangular-shaped configuration of 3 independent detectors co-located in Italy (ET-1, ET-2, ET-3) by using the 10 km arm ET-D noise curve model. While, for CE, we consider 2 independent L-shaped detectors: the first placed in the United States (CE1) and the second one in Australia (CE2), with 40 and 20 km arm lengths, respectively. 
In Fig.~\ref{fig: asd}, we depict the detector's amplitude spectral density (ASD) for the 2G (LVK)  and 3G detectors\footnote{The most recent ASDs of ET and CE can be found, respectively, at \url{https://www.et-gw.eu/index.php/etsensitivities} and \url{https://dcc.cosmicexplorer.org/CE-T2000017/public}.}. Furthermore, in Table \ref{tab:detectors}, we describe the main features of the interferometers: the localization, the orientation and the lowest frequency of the power spectral density\footnote{The locations and orientations of the interferometers are as reported in Table I of Ref.~\cite{Muttoni:2023prw}. }.
In the present analysis, we consider the following configurations:  ET, ET and CE1 (ET\,+\,CE1), and ET along with the two CE detectors (ET\,+\,CE1\,+\,CE2).

We model the quantity $\tilde{h}_\text{GR}$ in Eq.~\eqref{eq: ppE} with the \texttt{IMRPhenomD} waveform, considering an orbital configurations with spins aligned with the angular momentum.
Within this prescription, the set of binary parameters is $\mathcal{B}=\{\mathcal{M},\, q,\, d_L,\, \iota,\, t_c,\,\phi_c,\,\psi,\, \chi_1,\,\chi_2,\, ra,\, dec,\,\bm{\theta}_\text{PPE}\}$.
We can distinguish the extrinsic and intrinsic parameters. The former include the sky angles ($ra$, $dec$), the inclination $\iota$, the polarization angle $\psi$, the phase at coalescence $\phi_c$, the coalescence time $t_c$, and the luminosity distance of the source, $d_L$. On the other hand, the intrinsic parameters are the chirp mass $\mathcal{M}$, the mass ratio $q$, and the projection $\chi_i$ of the $i$-th spin along $z$.
Moreover, $\bm{\theta}_\text{PPE}$ represents the set of PPE expansion parameters encoding the parity-violating effects of the gravity scenarios under study. It is worth noticing that $\bm{\theta}_\text{PPE}$ is independent of the localization parameters ($ra$, $dec$, $\psi$, $\iota$).

To simulate the injection and to analyze the GW waveform, we adopt the open software \texttt{bilby} \cite{Ashton:2018jfp,Romero-Shaw:2020owr}.
The synthetic signal is taken into account by assuming the system parameters as in the event GW150914 \cite{LIGOScientific:2016aoc}. 
Furthermore, the PPE parameters are set to their corresponding GR fiducial values.
The fiducial values for the binary parameters are reported in Table \ref{tab: input value}. In Fig.~\ref{fig: strain}, we highlight the differences in the waveform when the $\bm{\theta}_\text{PPE}$ parameters are not vanishing.

Assuming the detector noise to be stochastic, stationary and a Gaussian function of time, we can evaluate the signal-to-noise ratio (SNR) through the expression
\begin{equation}
    \label{eq: SNR}
    \text{SNR} = \sqrt{\langle h,h\rangle}\,,
\end{equation}
where the inner product $\langle\cdot,\cdot\rangle$ is defined as
\begin{equation}
    \label{eq:inner prod}
    \langle A,B\rangle = 4\, \text{Re} \int_0^{\infty} \frac{\tilde{A}^{*}(f)\tilde{B}(f)}{S_n(f)}df\,,
\end{equation}
and $S_n(f)$ is the one-side power spectrum of the detector. For a network of $N$ detectors, the total SNR is given by
\begin{equation}
    \label{eq:snr_net} \text{SNR}_N=\sqrt{\sum_{i=1}^{N}\text{SNR}^{2}_{i}}\, .
\end{equation}
The estimated SNR values for the injected signal are 935, 1740 and 1811 for the ET, ET\,+\,CE1 and ET\,+\,CE1\,+\,CE2 networks, respectively.
Since the SNR is very high, we expect the localization parameters ($ra,\, dec,\, \psi,\, \iota$) to be weakly correlated with the intrinsic parameters. Hence, adopting the same approach as that used in the recent work \cite{Vaglio:2023lrd}, we fix the localization parameters to their fiducial values.
In so doing, the inference parameter set reduces to 
\begin{equation}
    \mathcal{B}=\{\mathcal{M}, q, d_L,\, t_c,\,\phi_c,\, \chi_1,\,\chi_2,\,\bm{\theta}_\text{PPE}\} .
\end{equation}

\begin{table}
    \setlength{\tabcolsep}{0.5em}
    \renewcommand{\arraystretch}{1.3}
    \begin{tabular}{c|c||c|c}
    \hline
       Parameter  & Value &Parameter  & Value \\
         \hline
         $\mathcal{M}\, [M_{\odot}]$& 28.1 &$\psi$ [rad]& 2.66\\
         $q$& 0.81&$ra$ [rad]& 1.38\\
         $d_L\, [\text{Mpc}]$& 400 &$dec$ [rad]& $-1.21$\\
         $\chi_1$& 0.31 &$\iota$ [rad]& 0.40\\
         $\chi_2$& 0.39 &$\phi_c$ [rad]& 1.30\\
          $t_c\, [\text s]$& 0.00& $\bm{\theta}_\text{PPE}$& 0.00\\
         \hline
    \end{tabular}
    \caption{Injection parameters for the binary system.}
    \label{tab: input value}
\end{table}

Therefore, we sample the posterior distributions by the \texttt{bilby-mcmc} algorithm \cite{Ashton:2021anp}, using the priors shown in Table~\ref{tab:priors}.
In our numerical analysis, we marginalize over the phase $\phi_c$ and coalescence time $t_c$, and we set the minimum frequency to $10$ Hz and the maximum frequency to $1024$ Hz  and the signal duration of $64$ s.
Additionally, we fix $H_0 = 67.7 $ km s$^{-1}$ Mpc$^{-1}$ and $\Omega_{m0}=0.308$ in order to convert the $d_L$ sampling into that over $z$. 
In what follows, we present the numerical constraints on the PPE parameters for the different theoretical scenarios.

\begin{table}
    \setlength{\tabcolsep}{0.5em}
    \renewcommand{\arraystretch}{1.3}
    \begin{tabular}{c|c}
    \hline
       Parameter  & Prior \\
       \hline
         $\mathcal{M}\, [M_{\odot}]$&$\mathcal{U}(20,100)$ \\
         $q$& $\mathcal{U}(0.125,1)$ \\
         $d_L\, [\text{Mpc}]$& $\mathcal{U}(100,5000)$  \\
         $\chi_1$&$\mathcal{U}(-1,1)$ \\
         $\chi_2$& $\mathcal{U}(-1,1)$\\
         $\bm{\theta}_\text{PPE}$ &$\mathcal{U}(-500,500)$ \\
         \hline
    \end{tabular}
    \caption{Priors for the free parameters of the sampling, where $\mathcal{U}$ indicates a uniform distribution function.}
    \label{tab:priors}
\end{table}

\subsection{Chern-Simons gravity}
\begin{figure*}
    \centering
    \includegraphics[width=0.95\textwidth]{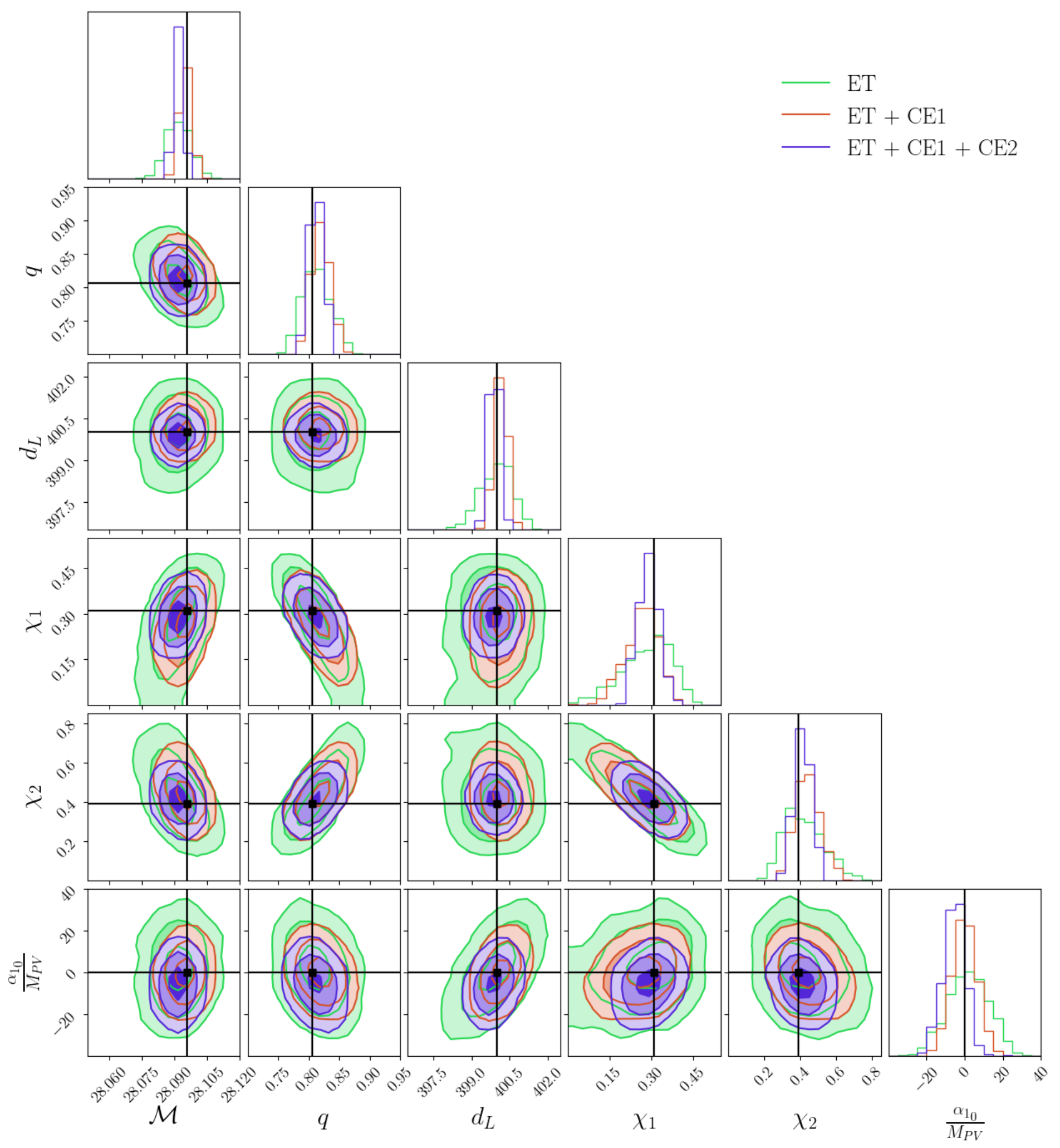}
    \caption{68\%, 95\% and 99\% C.L. contours, with posterior distributions, for the free parameters of CS gravity under different detector configurations. The straight lines indicate the injected values of the GW parameters. }
    \label{fig: corner CS}
\end{figure*}

From Eqs.~\eqref{eq: alpha cs} and \eqref{eq: beta cs}, the PPE parameter for CS gravity is
\begin{equation}
     \bm{\theta}_\text{PPE} =\frac{\alpha_{1_0}}{M_\text{PV}}\,.
\end{equation}
In Table~\ref{tab: CS results}, we present the results of our analysis for the different detector networks, whereas, in Fig.~\ref{fig: corner CS}, we show the $1\sigma$, $2\sigma$ and $3\sigma$ confidence level (C.L.) regions and the posterior distributions of the GW parameters.
In particular, we note that the PPE parameter is weakly correlated with $d_L$ and $\chi_1$.
The PPE parameter is constrained with an accuracy of $(10.93,\, 6.70,\, 5.91)\, M_{\odot} $ for ET, ET\,+\,CE1 and ET\,+\,CE1\,+\,CE2, respectively.

\begin{figure*}
    \centering
    \includegraphics[width=0.9\textwidth]{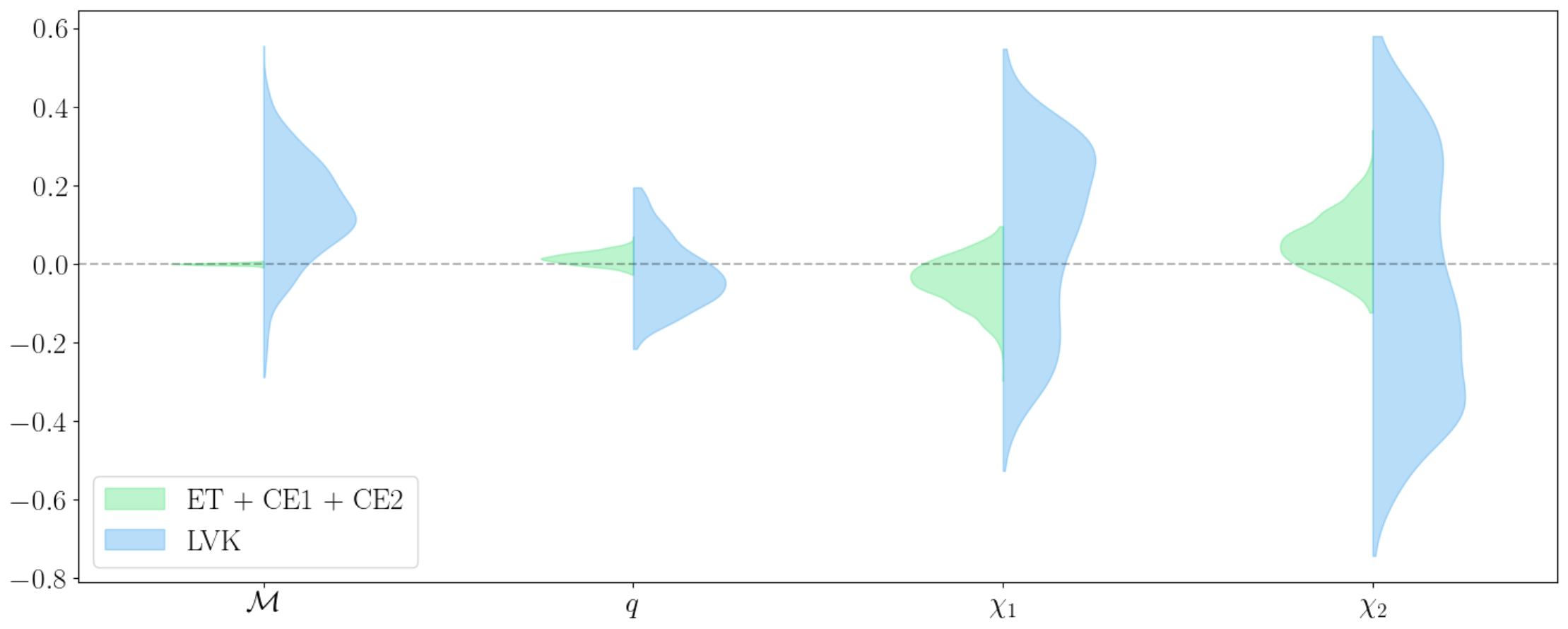}\\
    \includegraphics[width=0.9\textwidth]{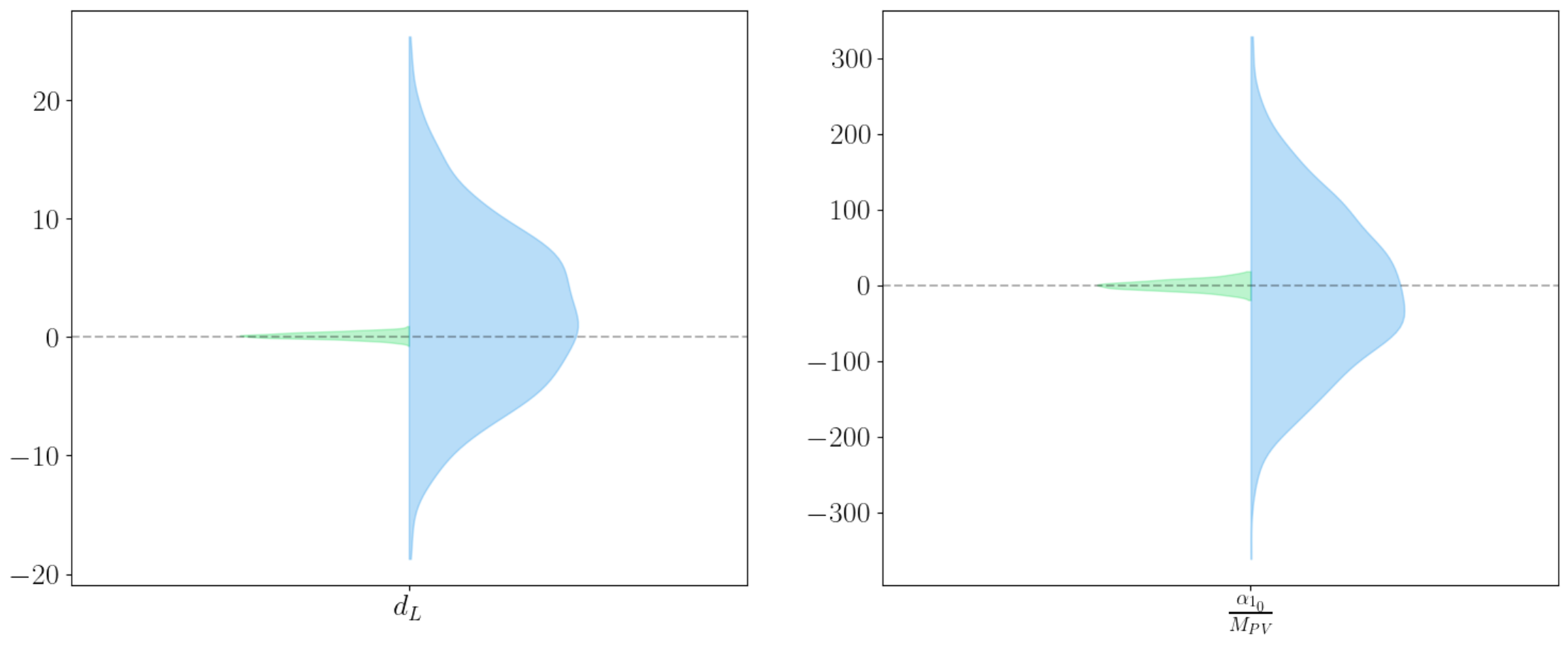}
    \caption{Deviations of the posterior probability distributions from the injected values of the GW parameters for CS gravity. The injected signal is analyzed both for the 3G detector (ET\,+\,CE1+\,CE2) and 2G detector (LVK) configurations. 
    The horizontal dashed lines indicate zero deviations from the injected signal.}
    \label{fig: corner CS LVK}
\end{figure*}

In Fig.~\ref{fig: corner CS LVK}, we compare the results obtained from 3G detectors with those of the 2G detector network, keeping the localization parameters fixed at their fiducial values.
Specifically, we quantify the deviations of the posterior distributions from the injected values of the GW parameters.
As such, we highlight an improvement on the PPE parameter of a factor $\sim 18$.

\subsection{Symmetric teleparallel gravity}
\begin{figure*}
    \centering
    \includegraphics[width=1\textwidth]{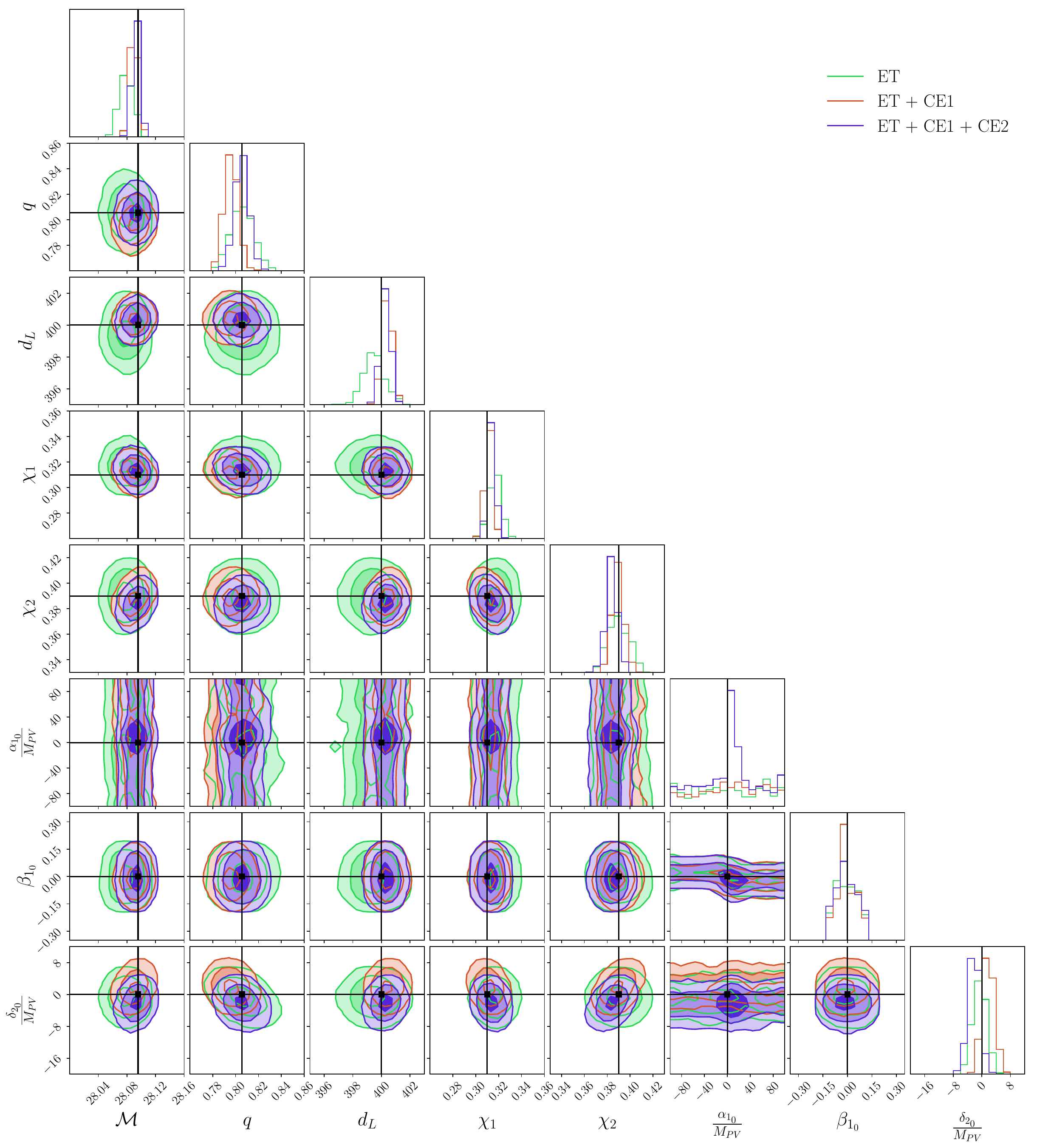}
    \caption{68\%, 95\% and 99\% C.L. contours, with posterior distributions, for the free parameters of ST gravity under different detector configurations. The straight lines indicate the injected values of the GW parameters.
    }
    \label{fig: corner fQ}
\end{figure*}

Given Eqs.~\eqref{eq: a 1 ppe stegr} to \eqref{eq: b 2 ppe stegr}, we can define the PPE parameter set in ST gravity  as follows:
\begin{equation}
    \bm{\theta}_\text{PPE} = \left\{ \frac{\alpha_{1_0}}{M_\text{PV}},\, \beta_{1_0},\, \frac{\delta_{2_0}}{M_\text{PV}}\right\}.
\end{equation}
The MCMC results are listed in Table~\ref{tab: STEGR results} and plotted in Fig.~\ref{fig: corner fQ}. 
It is worth noticing that, in all configurations the quantity $\frac{\alpha_{1_0}}{M_\text{PV}}$ turns out to be unconstrained, as
as it is not characterized by a specific posterior distribution, which simply reflects the chosen priors. The same behavior occurs also by enlarging the priors.
On the other hand, the parameter $\beta_{1_0}$ is bounded with an accuracy of 0.08, 0.06 and 0.074 under the ET, ET\,+\,CE1 and ET\,+\,CE1\,+\,CE2 configurations, respectively. The same detector networks are capable of constraining $\frac{\delta_{2_0}}{M_\text{PV}}$ with an accuracy of $(1.80,\, 1.70,\, 1.53)\, M_{\odot}^2 \,\text{Mpc}^{-1}$, respectively.

\begin{figure*}
    \centering
    \includegraphics[width=0.9\textwidth]{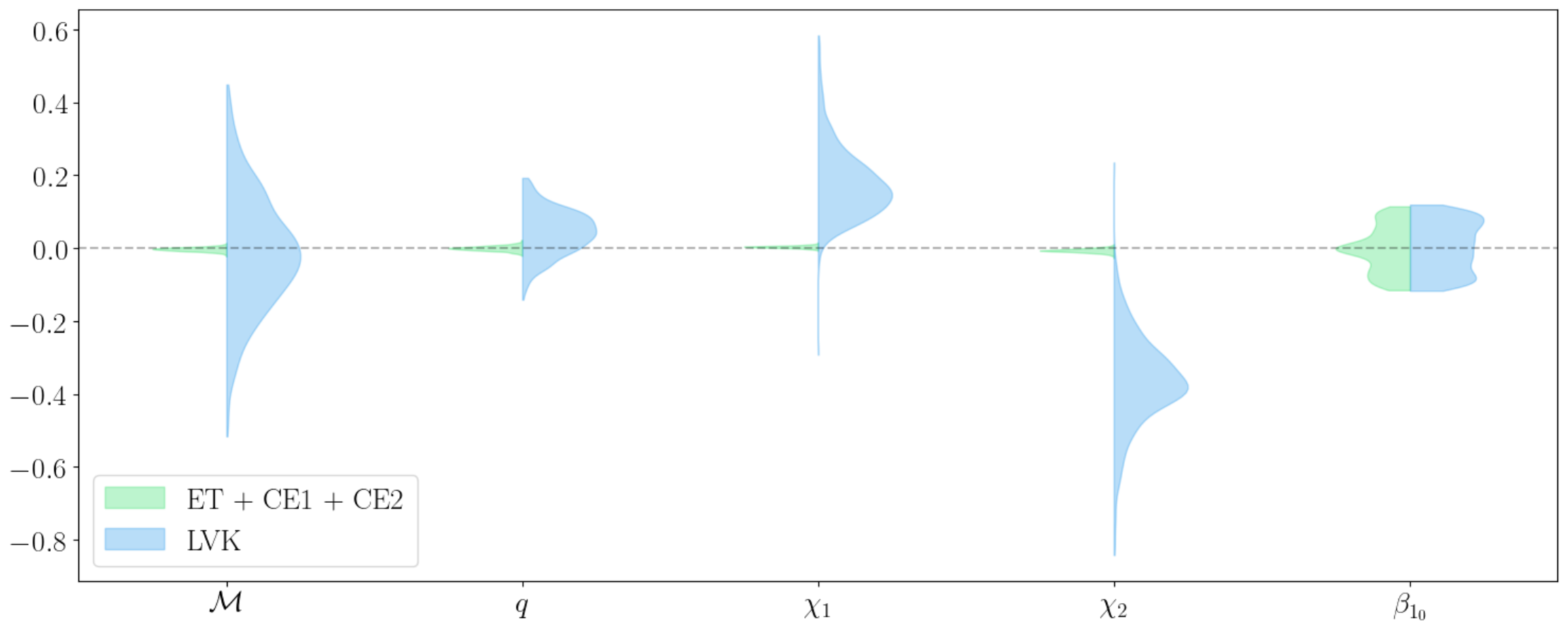}\\
    \includegraphics[width=0.9\textwidth]{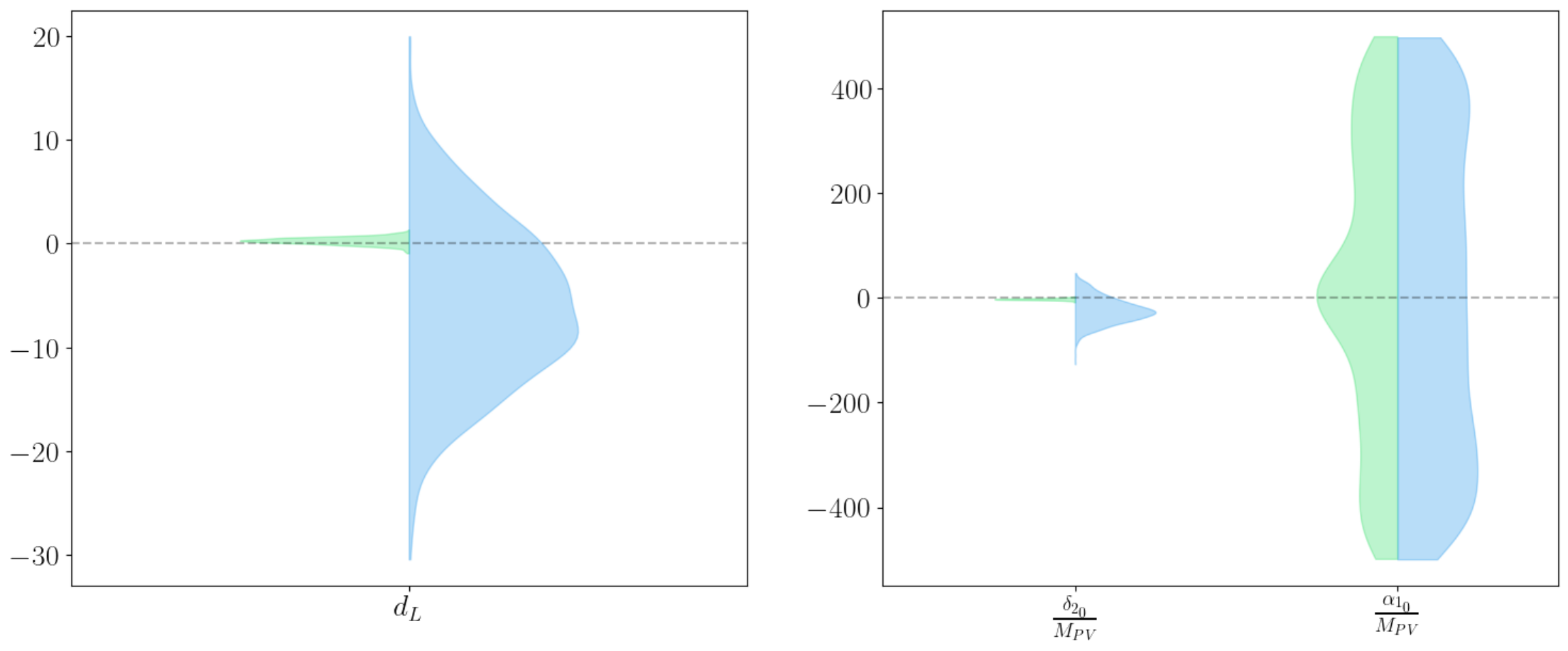}
    \caption{Deviations of the posterior probability distributions from the injected values of the GW parameters for ST gravity. The injected signal is analyzed both for the 3G detector (ET\,+\,CE1+\,CE2) and 2G detector (LVK) configurations. 
    The horizontal dashed lines indicate zero deviations from the injected signal.}
    \label{fig: corner fQ LVK}
\end{figure*}

Similarly to the case of CS gravity, we compare the results obtained for the 3G and 2G detector networks in Fig.~\ref{fig: corner fQ LVK}.
We note that $\frac{\alpha_{1_0}}{M_\text{PV}}$ remains unconstrained also for the 2G detectors.
Moreover, the two configurations provide similar accuracy on the parameter $\beta_{1_0}$. However, the posterior distribution of the latter from the 3G detectors peaks around $0$, while the result of the 2G detectors is almost flat in the same confidence interval.
Finally, the 3G detector network improves the accuracy on $\frac{\delta_{2_0}}{M_\text{PV}}$ by a factor $\sim 15$.

\subsection{Ho\v rava-Lifshitz gravity}
\begin{figure*}
    \centering
    \includegraphics[width=0.98\textwidth]{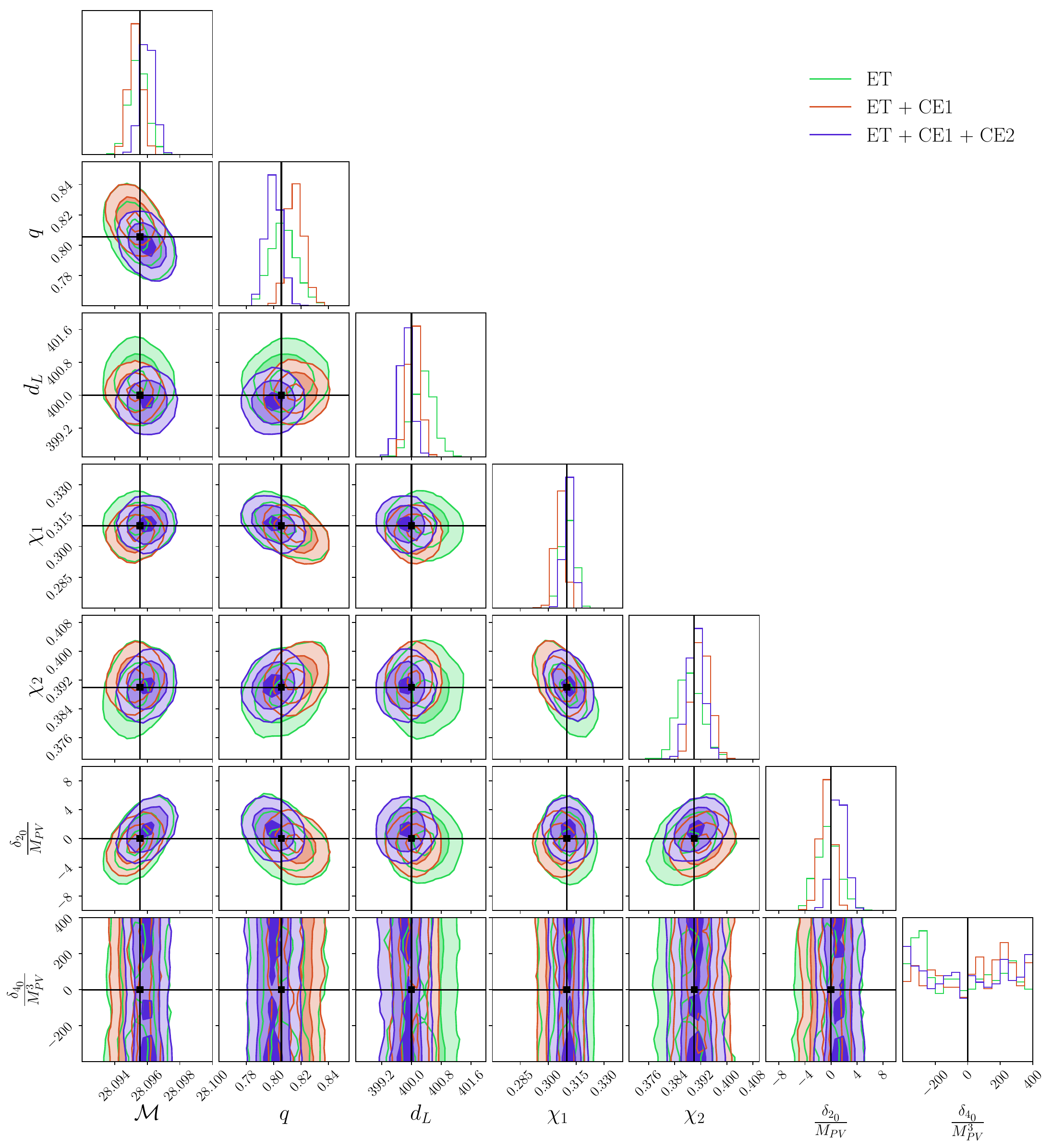}
    \caption{68\%, 95\% and 99\% C.L. contours, with posterior distributions, for the free parameters of HL gravity under different detector configurations. The straight lines indicate the injected values of the GW parameters.
   }
    \label{fig: corner HL}
\end{figure*}

The PPE parameter set in the case of HL gravity is provided by Eqs.~\eqref{eq: a 2 HL} to \eqref{eq: b 4 HL}:
\begin{equation}
    \bm{\theta}_\text{PPE} = \left\{ \frac{\delta_{2_0}}{M_\text{PV}},\,  \frac{\delta_{4_0}}{M_\text{PV}^3}\right\}.
\end{equation}
We show the posterior distributions in Fig.~\ref{fig: corner HL}, and the best-fit values of the GW parameters in Table~\ref{tab: HL results}. We can see that $\frac{\delta_{4_0}}{M_\text{PV}^3}$ is unconstrained, while $\frac{\delta_{2_0}}{M_\text{PV}}$ is bounded with an accuracy of 1.78, 1.08 and  1.00 under the ET, ET\,+\,CE1 and ET\,+\,CE1\,+\,CE2 networks, respectively.

\begin{figure*}
    \centering
    \includegraphics[width=0.9\textwidth]{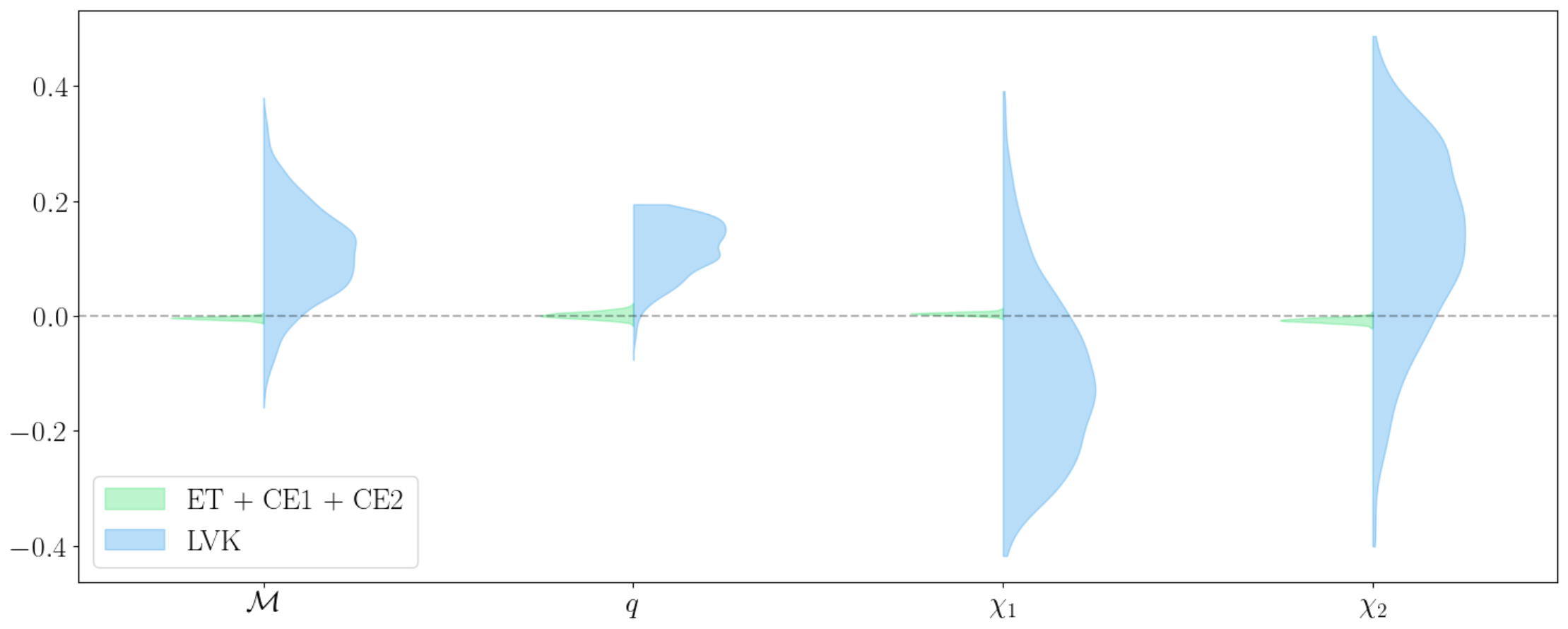}\\
    \includegraphics[width=0.9\textwidth]{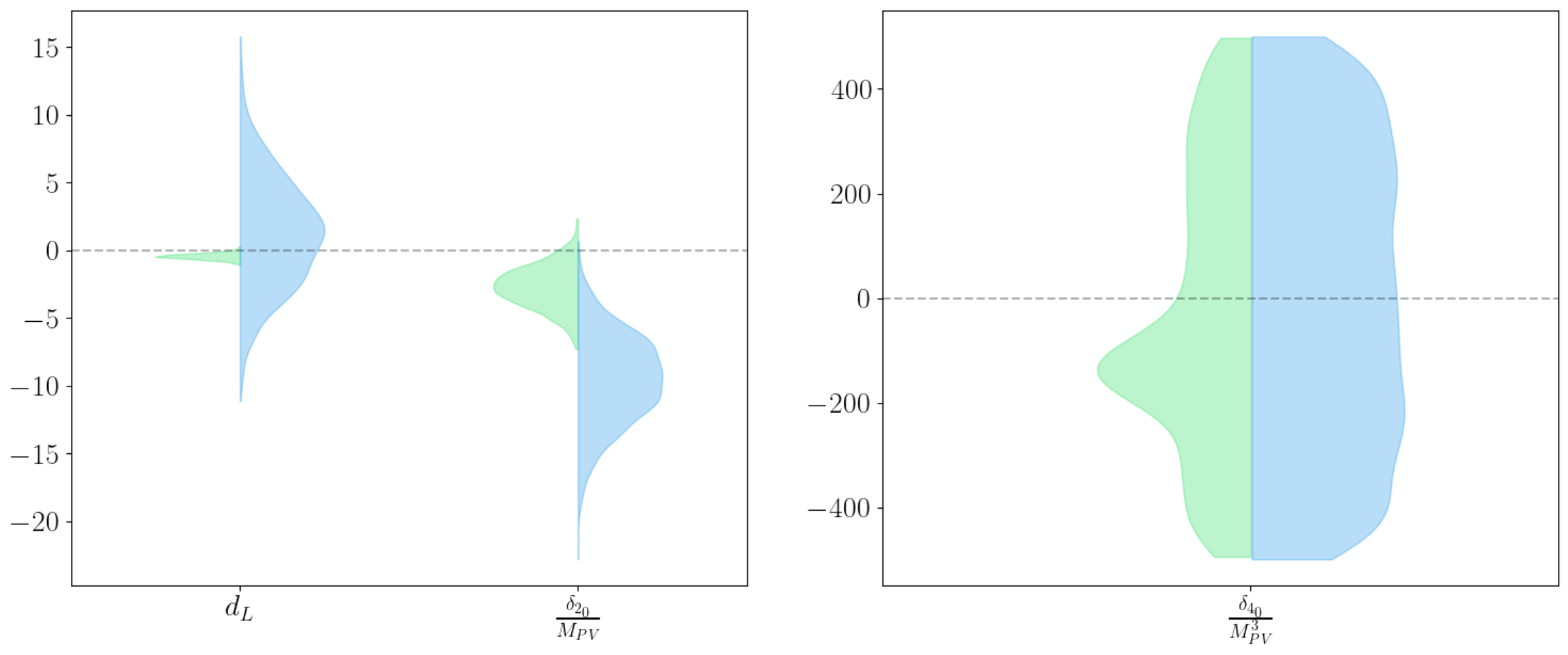}
    \caption{Deviations of the posterior probability distributions from the injected values of the GW parameters for HL gravity. The injected signal is analyzed both for the 3G detector (ET\,+\,CE1+\,CE2) and 2G detector (LVK) configurations. 
    The horizontal dashed lines indicate zero deviations from the injected signal.}
    \label{fig: corner HL LVK}
\end{figure*}

Furthermore, also for HL gravity, in Fig.~\ref{fig: corner HL LVK} we highlight the improvement one may obtain through 3G detectors compared to the 2G detector network. In fact, the accuracy on  $\frac{\delta_{2_0}}{M_\text{PV}}$ increase by a factor $\sim 19$.

\begin{table*}
    \subfloat[CS gravity]
    {
    \setlength{\tabcolsep}{0.7em}
    \renewcommand{\arraystretch}{1.8}
    \begin{tabular}{c|cccccc}
  	\hline
    Network & $\mathcal{M}\, [M_\odot]$ & $q$ & $d_L\, [\text{Mpc}]$ & $\chi_1$ & $\chi_2$ & $\frac{\alpha_{1_0}}{M_\text{PV}}\, [M_\odot]$ \\
    \hline
    ET & $28.09^{+0.01}_{-0.01}$ & $0.81^{+0.03}_{-0.02}$ & $400.02^{+0.58}_{-0.67}$ & $0.29^{+0.08}_{-0.11}$ & $0.41^{+0.12}_{-0.10}$ &$1.82^{+11.04}_{-10.83}$ \\
    ET\,+\,CE1 &  $28.092^{+0.003}_{-0.003}$ & $0.81^{+0.02}_{-0.02}$ & $400.14^{+0.27}_{-0.25}$ & $0.27^{+0.05}_{-0.07}$ & $0.44^{+0.08}_{-0.07}$ &$-0.28^{+6.91}_{-6.49}$\\
    ET\,+\,CE1\,+\,CE2 & $28.092^{+0.002}_{-0.002}$ & $0.81^{+0.01}_{-0.01}$ & $399.88^{+0.20}_{-0.22}$ & $0.29^{+0.04}_{-0.04}$ & $0.41^{+0.05}_{-0.05}$ &$-5.27^{+5.47}_{-6.35}$ \\
    \hline
    \end{tabular}
     \label{tab: CS results}
    }
    \\
    \subfloat[ST gravity]
    {
    \setlength{\tabcolsep}{0.2em}
    \renewcommand{\arraystretch}{1.8}
    \begin{tabular}{c|cccccccc}
    \hline
    Network &$\mathcal{M}\, [M_{\odot}]$  & $q$ & $d_L\, [\text{Mpc}]$ & $\chi_1$ & $\chi_2$ &$\frac{\alpha_{1_0}}{M_\text{PV}}\,[M_\odot]$ & $\beta_{1_0}\, [\text{Mpc}^{-1}]$ & $\frac{\delta_{2_0}} {M_\text{PV}}\, [M_{\odot}^2 \,\text{Mpc}^{-1}]$\\
    \hline
    ET&$28.10^{+0.01}_{-0.01}$ & $0.81^{+0.01}_{-0.01}$ & $399.50^{+0.38}_{-0.38}$ & $0.32^{+0.01}_{-0.01}$ & $0.39^{+0.01}_{-0.01}$ & \emph{n.c.} &$0.00^{+0.08}_{-0.08}$  & $-1.34^{+1.81}_{-1.78}$\\
    ET\,+\,CE1 & $28.090^{+0.005}_{-0.005}$ & $0.796^{+0.006}_{-0.006}$ & $400.39^{+0.36}_{-0.34}$ & $0.312^{+0.003}_{-0.003}$ & $0.389^{+0.005}_{-0.005}$ & \emph{n.c.} &$-0.03^{+0.09}_{-0.03}$  & $1.79^{+1.78}_{-1.63}$\\
    ET\,+\,CE1\,+\,CE2 & $28.093^{+0.005}_{-0.005}$ & $0.805^{+0.006}_{-0.006}$ & $400.26^{+0.32}_{-0.33}$ & $0.314^{+0.003}_{-0.003}$ & $0.384^{+0.005}_{-0.005}$ & \emph{n.c.}&$-0.002^{+0.074}_{-0.074}$  & $-2.19^{+1.55}_{-1.52}$\\
   \hline
   \end{tabular}
	\label{tab: STEGR results}
	}
    \\
    \subfloat[HL gravity]
    {
    \setlength{\tabcolsep}{0.4em}
    \renewcommand{\arraystretch}{1.8}
    \begin{tabular}{c|ccccccc}
   	\hline
    Network &$\mathcal{M}\, [M_{\odot}]$  & $q$ & $d_L\, [\text{Mpc}]$ & $\chi_1$ & $\chi_2$ &$\frac{\delta_{2_0}}{M_\text{PV}}\, [M_{\odot}^2 \,\text{Mpc}^{-1}]$ & $ \frac{\delta_{4_0}}    {M_\text{PV}^3}\, [M_{\odot}^4 \,\text{Mpc}^{-1}]$\\
    \hline
     ET&$28.08^{+0.01}_{-0.01}$ & $0.81^{+0.01}_{-0.01}$ & $400.33^{+0.33}_{-0.30}$ & $0.31^{+0.01}_{-0.01}$ & $0.39^{+0.01}_{-0.01}$ & $-0.44^{+1.72}_{-1.84}$& \emph{n.c.}  \\
     ET\,+\,CE1 &$28.100^{+0.007}_{-0.007}$ & $0.80^{+0.01}_{-0.01}$ & $399.86^{+0.17}_{-0.17}$ & $0.306^{+0.004}_{-0.004}$ & $0.393^{+0.006}_{-0.006}$ & $1.23^{+1.12}_{-1.03}$& \emph{n.c.}  \\
     ET\,+\,CE1\,+\,CE2 &$28.092^{+0.003}_{-0.003}$ & $0.82^{+0.01}_{-0.01}$ & $400.07^{+0.16}_{-0.17}$ & $0.314^{+0.003}_{-0.003}$ & $0.383^{+0.004}_{-0.004}$ & $-0.72^{+1.00}_{-0.99}$& \emph{n.c.}  \\
         \hline
    \end{tabular}
    \label{tab: HL results}	
            }
        \caption{Best-fit values and 1$\sigma$ uncertainties on the GW parameters of the theoretical scenarios under study, for different detector configurations. \emph{n.c.} stands for \emph{not 		constrained.}}
\end{table*}

\section{Summary and discussion}
\label{sec:conclusions}

We considered parity violation in the propagation of GWs through a newly proposed PPE formalism. In particular, we framed deviations from GR through a general parametrized framework taking into account the modified amplitude and phase of GWs. We focused on the cases of CS, ST and HL gravity, where departures from Einstein's theory may emerge from additional parity-violating terms included in the gravitational action. Then, we outlined the geometrical and physical characteristics of future ground-based GW interferometers, such as ET and CE. 
We showed how they can be used to probe parity violations, and we described the method to constrain the PPE expansion parameters.

Using the sensitivities of 3G detectors, we simulated GW signals from binary systems, such as BBH and BNS, and we obtained $68\%$, $95\%$ and $99\%$ numerical bounds on both binary and PPE parameters, for different GW detector networks.
The accuracy of the GW and PPE parameters increases when more detectors are considered in the network, independently from the theoretical framework. As the SNR is very high, the uncertainties on the GW parameters turn out to be quite low. Indeed, for all models under study, we constrained the chirp mass and mass ratio with a relative accuracy of $\sim (0.04,\, 0.03,\, 0.01) \%$ and $\sim(3,\, 1.5,\,1)\%$ for ET, ET\,+\,CE1 and ET\,+\,CE1\,+\,CE2, respectively. 
Additionally, the precision on $d_L$ spans from $\sim 0.2\%$ in the case of ET alone, to $\sim 0.1\% $ and $\sim 0.08 \%$ when ET is combined with one or two CE detectors.
Moreover, we bounded the spin parameter $\chi_1$ with a relative accuracy of $\sim (20,\, 10,\, 3 )\%$ and $\sim(20,\,10,\,2)\%$ for the three detector configurations, respectively. 
As regards the PPE parameters, we found that one of them remains unconstrained in ST and HL gravity.
This feature may be related to the low-frequency cut-off. In fact, to reduce the computational time, we fixed the minimum frequency of $10$ Hz. However, one might extend the analysis to the 1-10 Hz frequency band, and increase the duration of the signal, to improve the constraints on GW parameters. 

Furthermore, we compared the results of the combined 3G detectors with those of the 2G detector configuration of LVK interferometers. For each parity-violating model, we showed the deviations of the posterior distributions of the fitting parameters with respect to the injected GR signal.
Our results indicate an improvement of roughly one order of magnitude compared to those obtained for the 2G detectors. Specifically, from the LVK configuration, we obtained a relative accuracy of $0.5\%$ on $\mathcal{M}$, $7 \% $ on the mass ratio, $1.6 \% $ on $d_L$, and $63 \% $ and $80 \% $ on $\chi_1$ and $\chi_2$, respectively.
We note that the constraint on the $\tilde{h}_\text{GR}$ waveform parameters are almost independent of the theoretical model also under the LVK analysis.

Finally, it is worth to stress that the PPE parameters enter the waveform at higher orders of the PN expansion. Hence, a more accurate waveform would be needed in the future to detect with greater precision deviations from GR that arise from parity-violating theories.
In future investigations, we also plan to perform a more comprehensive Bayesian analysis, allowing the localization parameters to vary freely in the numerical simulations.

\acknowledgments

The authors acknowledge the financial support of INFN - Sezione di Napoli, {\it iniziative specifiche} QGSKY, MOONLIGHT and TEONGRAV. 
D.V. acknowledges the FCT Project No. PTDC/FIS-AST/0054/2021.

\clearpage

\bibliography{references}

\end{document}